\documentclass[prd,aps,floats,floatfix,eqsecnum,nofootinbib]{revtex4}
\usepackage{amsmath,amssymb,verbatim,epsfig,graphicx,rotating}
\usepackage{psfrag}
\usepackage{multirow}
\newcommand{\be}{\begin{equation}}
\newcommand{\ee}{\end{equation}}
\newcommand{\bea}{\begin{eqnarray}}
\newcommand{\eea}{\end{eqnarray}}
\newcommand{\vx}{\ensuremath{\vec{x}}}

\begin{document}
\title{The Effective Theory of Inflation and the Dark Matter Status
in the Standard Model of the Universe}
\author{\bf H. J. de Vega $^{(a,b)}$} \email{devega@lpthe.jussieu.fr}
\affiliation{$^{(a)}$ LPTHE, Laboratoire Associ\'e au CNRS UMR 7589,\\
Universit\'e Pierre et Marie Curie (Paris VI) et Denis Diderot 
(Paris VII), \\ Tour 24, 5 \`eme. \'etage, 4, Place Jussieu, 
75252 Paris, Cedex 05, France.\\ 
$^{(b)}$ Observatoire de Paris, LERMA, Laboratoire Associ\'e au 
CNRS UMR 8112, \\61, Avenue de l'Observatoire, 75014 Paris, France}

\date{\today}

\begin{abstract}
Inflation is today a part of the Standard Model of the Universe supported by
the cosmic microwave background (CMB) and large scale structure (LSS) datasets.
Inflation solves the horizon and flatness problems and 
naturally generates  density fluctuations that seed LSS and CMB anisotropies, 
and tensor perturbations (primordial gravitational waves). Inflation theory is 
based on a scalar field  $ \varphi $ (the inflaton) whose potential 
is fairly flat leading to a slow-roll evolution. We present here the
effective theory of inflation \`a la Ginsburg-Landau in which
the inflaton potential is a polynomial in the field $ \varphi $ and has
the universal form $ V(\varphi) = N \; M^4 \;
w(\varphi/[\sqrt{N}\; M_{Pl}]) $, where $ w = {\cal O}(1) , \;
M \ll M_{Pl} $ is the scale of inflation and  $ N \sim 60 $ is the number 
of efolds since the cosmologically relevant modes exit the horizon till 
inflation ends.
The slow-roll expansion becomes a systematic $ 1/N $ expansion and
the inflaton couplings become {\bf naturally small} as powers of the ratio
$ (M / M_{Pl})^2 $. The spectral index and the ratio of tensor/scalar
fluctuations are $ n_s - 1 = {\cal O}(1/N), \; r = {\cal O}(1/N) $ while
the running index turns to be $ d n_s/d \ln k =  {\cal O}(1/N^2) $
and therefore can be neglected. The energy scale of inflation $ M \sim 0.7 
\times 10^{16}$ GeV is
completely determined by the amplitude of the scalar adiabatic
fluctuations. A complete analytic study plus the 
Monte Carlo Markov Chains (MCMC) analysis of the available
CMB+LSS data (including WMAP5) with fourth degree potentials 
showed: (a) the {\bf spontaneous breaking} of the
$ \varphi \to - \varphi $ symmetry of the inflaton potential. (b)
a {\bf lower bound} for $ r $ in new inflation:
$ r > 0.023 \; (95\% \; {\rm CL}) $ and $ r > 0.046 \;  (68\% \;
{\rm CL}) $. (c) The preferred inflation potential is a {\bf double
well}, even function of the field with a moderate quartic coupling
yielding as most probable values: $ n_s \simeq 0.964 , \; r\simeq
0.051 $. This value for $ r $ is within reach of forthcoming CMB
observations. Study of higher degree inflaton potentials show that terms 
of degree higher than four do not affect the fit in a significant way. 
The initial conditions for the quantum fluctuations must be
vacuum type (Bunch-Davies) in order to reproduce the CMB and LSS data.
Slow-roll inflation is generically preceded by a short {\bf fast-roll} stage. 
If the modes which are horizon-size today exited the horizon during 
fast-roll or at the transition between fast and slow-roll, the 
curvature and tensor CMB quadrupoles get suppressed in agreement with
the CMB data for the former.  Fast-roll fits the TT, the TE and the EE
modes well reproducing the quadrupole supression. A thorough study of the 
quantum loop corrections reveals that they are very small and controlled by
powers of $(H /M_{Pl})^2 \sim {10}^{-9} $, a conclusion that validates the 
reliability of the effective theory of inflation. Our work shows how powerful is
the Ginsburg-Landau effective theory of inflation in predicting
observables that are being or will soon be contrasted to observations.
Dark matter (DM) constitutes 83 \% of the matter in the Universe.
We investigate the DM properties using cosmological theory and the
galaxy observations from DM-dominated galaxies.
Our DM analysis is independent of the particle physics
model for DM and it is based  on the DM phase-space density $ \rho_{DM}/\sigma^3_{DM} $.
We derive explicit formulas for the DM particle
mass $ m $ and for the number of ultrarelativistic degrees of freedom $ g_d $
(hence the temperature) at decoupling. We find that $ m $ turns to be at the 
{\bf keV scale}. The keV scale DM is non-relativistic during structure formation,
reproduces the small and large scale structure but it cannot be
responsible of the $ e^+ $ and $ \bar p $ excess in cosmic rays which can be 
explained by astrophysical mechanisms.
\end{abstract}

\maketitle

\tableofcontents

\section{Overview and present status of  the Effective Theory of
Inflation}

This article is dedicated to my colleague and friend Lev Lipatov
in the ocassion of his 70th birthdate. The content of this
paper is not directly related to the monumental work of Lev in 
particle physics and field theory. However, the deep subtleties in
the field theoretical treatments in cosmology 
desserve to include this contribution in Lev's Festschrift.

\bigskip

Inflation was introduced to solve several outstanding problems of the
standard Big Bang model \cite{guthsato} and has now become an essential 
part of the standard cosmology. It provides a natural mechanism
for the generation of scalar density fluctuations that seed large scale
structure, thus explaining the origin of the temperature anisotropies in
the cosmic microwave background (CMB), and for the generation of tensor
perturbations (primordial gravitational waves) 
\cite{kt,libros,hu,nos}.

\medskip

\begin{figure}[h]
\includegraphics[height=9.cm,width=12.cm]{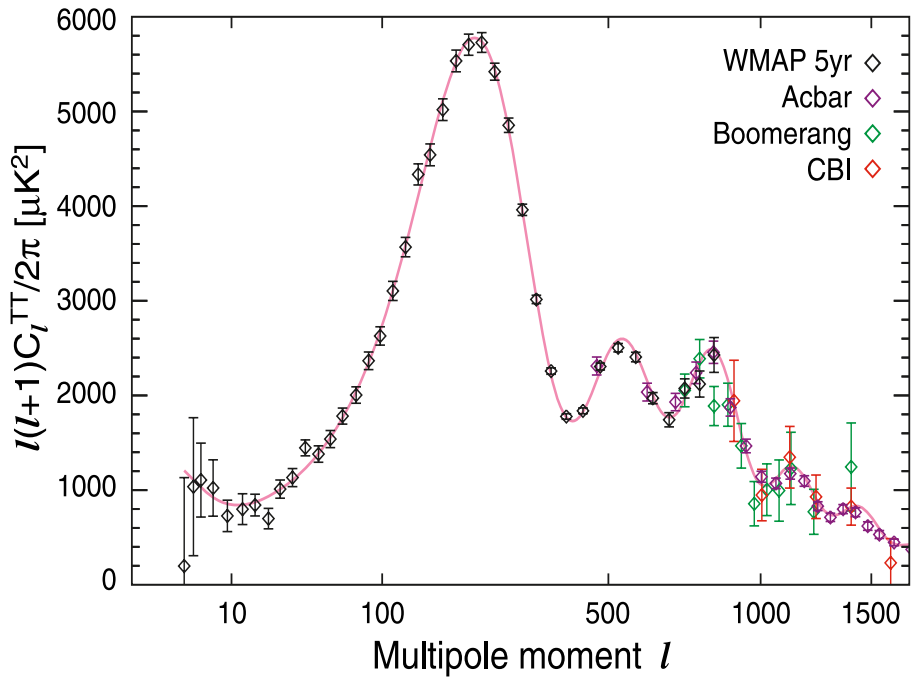}
\caption{Acoustic oscillations from WMAP 5 years data set plus other 
CMB data. Theory and observations nicely agree except for the lowest 
multipoles: the quadrupole CMB suppression. See refs. 
\cite{quadru1,quadru2,quamc} and \cite{nos} for discussions on this.} 
\label{acus}
\end{figure}

A distinct aspect of inflationary perturbations is that they are 
generated by quantum fluctuations of the scalar field(s) that drive 
inflation. After their wavelength becomes larger than the Hubble radius, 
these fluctuations are amplified and grow, becoming classical and 
decoupling from causal microphysical processes. Upon re-entering the 
horizon, during the radiation and matter dominated eras, these classical 
perturbations seed the inhomogeneities which generate structure upon 
gravitational collapse \cite{libros,hu}. 
These fluctuations enjoy fairly generic features: a
gaussian, nearly scale invariant spectrum of adiabatic scalar 
and tensor primordial fluctuations, which provide an
excellent fit to the highly precise wealth of CMB \cite{WMAP5} and LSS data,
making the inflationary paradigm robust.
Precision CMB data reveal peaks and valleys in the temperature
fluctuations resulting from acoustic oscillations in the
electron-photon fluid at recombination. These are depicted in fig. 
\ref{acus} where up to five peaks can be seen.

Baryon acoustic oscillations driven by primordial fluctuations
produce a peak in the galaxy correlations at 
$ \sim 109 \; h^{-1} $ Mpc (comoving
sound horizon) \cite{eis}. This peak is the real-space version of the 
acoustic oscillations in momentum (or $l$) space and are confirmed 
by LSS data \cite{eis}.

\begin{figure}[h]
\includegraphics[height=7.cm,width=12.cm]{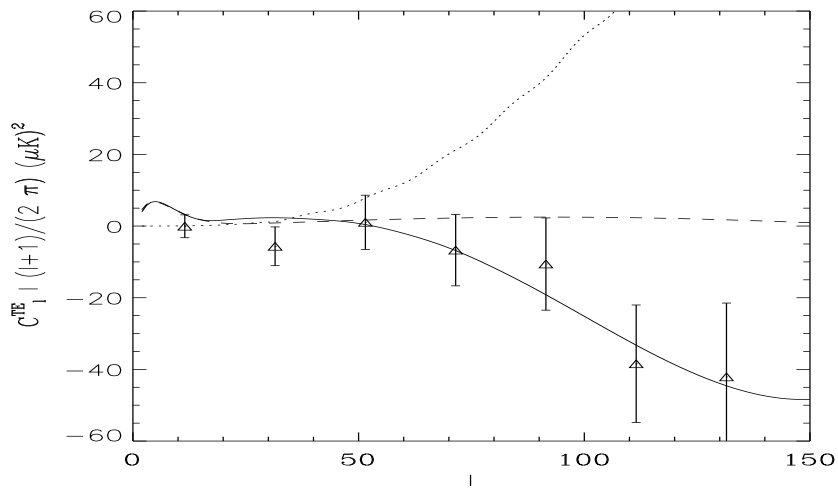}
\caption{Temperature-Polarization angular power spectrum.
The large-angle TE power spectrum predicted in primordial adiabatic 
models (solid), primordial isocurvature models (dashed) and by defects
such as cosmic strings (dotted). The WMAP TE data (Kogut et al. \cite{WMAP1})
are shown for comparison, in bins of $ \Delta l = 10 $. Superhorizon 
adiabatic modes from inflation fit the data while subhorizon sources of 
TE power go in directions opposite to the data. Hence, we concentrate here and
in ref. \cite{nos} on superhorizon adiabatic modes.}
\label{TE}
\end{figure}

Perhaps the most striking validation of inflation
as a mechanism for generating \emph{superhorizon} fluctuations
is the anticorrelation peak in the temperature-polarization (TE) angular
power spectrum at $ l \sim 150 $ corresponding to superhorizon
scales \cite{WMAP1} and depicted in fig. \ref{TE}. 
The observed TE power spectrum can only 
be generated by fluctuations that exited the horizon during inflation
and re-entered the horizon later, when the expansion of the universe
decelerates.

\begin{figure}[h]
\includegraphics[height=7.cm,width=12.cm]{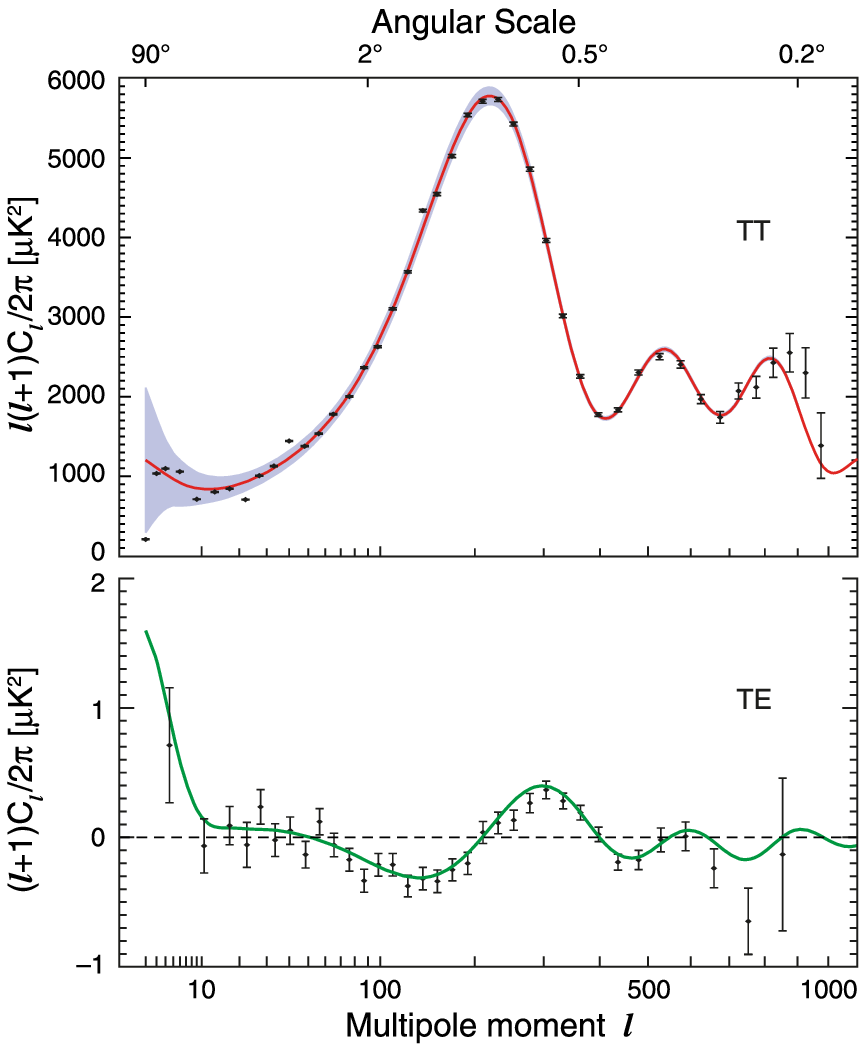}
\caption{The temperature (TT) and temperature-polarization correlation 
(TE) power spectra based on the 5 year WMAP data \cite{WMAP5}.}
\label{TEwmap5}
\end{figure}

The confirmation of many of the robust predictions of
inflation by current high precision observations places inflationary
cosmology on solid grounds.

\medskip

Amongst the wide variety of inflationary scenarios, single field slow-roll
models provide an appealing, simple and fairly generic description of
inflation. Its implementation is based on a scalar field (the
inflaton) whose homogeneous expectation value drives the dynamics of the
scale factor, plus small quantum fluctuations. The inflaton potential is
fairly flat during inflation and it dominates the universe energy during 
inflation. This flatness not only leads to a slowly varying Hubble 
parameter, hence ensuring a sufficient number of efolds of inflation,
but also provides an explanation for the gaussianity of the fluctuations as
well as for the (almost) scale invariance of their power spectrum. A flat
potential precludes large non-linearities in the dynamics of the
\emph{fluctuations} of the scalar field.

\medskip

The current WMAP data are validating the single field slow-roll scenario 
\cite{WMAP1,WMAP5}. Furthermore, because the potential is flat the
scalar field is almost {\bf massless}, and modes cross the horizon with an
amplitude proportional to the Hubble parameter. This fact combined with a
slowly varying Hubble parameter yields an almost scale invariant primordial
power spectrum.  The slow-roll approximation has been cast as a 
systematic $ 1/N $ expansion \cite{1sN}, where $ N \sim 60 $ is the number 
of efolds since the cosmologically relevant modes exited the horizon till 
 the end of inflation.

\medskip

The observational progress discriminates among 
different inflationary models, placing stringent constraints on them. 
The upper bound on the ratio $ r $ of tensor to scalar fluctuations 
obtained by WMAP convincingly excludes the massless monomial 
$ \varphi^4 $ potential \cite{WMAP5} and hence 
{\bf strongly suggests} the presence of a {\bf mass
term} in the single field inflaton potential \cite{ciri,infwmap,nos}.
Therefore, as a minimal single field model, 
one should consider a sufficiently 
general polynomial, the simplest polynomial potential bounded
from below being the fourth degree potential \cite{mcmc,ciri,nos}. 

\medskip

The observed low value of the CMB quadrupole with respect to
the $\Lambda$CDM theoretical value
has been an intriguing feature on large angular scales since 
first observed by COBE/DMR \cite{cobe}, and confirmed by the WMAP data 
\cite{WMAP5}.  In the best fit $\Lambda$CDM model, 
using the WMAP5 data we find that the probability that the quadrupole is as
low or lower than the observed value is just 0.031 \cite{nos,singu}. 
Even if one does not 
care about the specific multipole and looks for any multipole as low or 
lower than the observed quadrupole with respect to the $\Lambda$CDM model 
value, then the probability remains smaller than 5\%. Therefore, it is 
relevant to find a cosmological explanation of the quadrupole supression 
beyond the $\Lambda$CDM model. An early fast-roll stage can explain the CMB
 quadrupole suppression as we discuss below.

\medskip

The main new aspects of inflationary cosmology can be summarized as follows \cite{nos}:

\begin{itemize}
\item{An effective field theory description of slow-roll single
field inflation \`a la Ginsburg-Landau. In the Ginsburg-Landau framework,
the potential is a polynomial in the field starting by a constant term 
\cite{gl}. Linear terms can always be eliminated by a constant shift of 
the inflaton field. The quadratic term can have a positive or a
negative sign associated to chaotic or new inflation, respectively.
This effective Ginsburg-Landau field theory is characterized by only 
{\bf two} energy scales: the scale of inflation $ M $ and the Planck scale 
$ M_{Pl} = 2.43534 \; 10^{18}$ GeV $ \gg M $. In this context we propose a 
{\bf universal} form for the inflaton potential in slow-roll models \cite{1sN}:
\be \label{VIn}
V(\varphi) = N \; M^4 \; w(\chi)  \; ,
\ee
\noindent where $ N \sim 60 $  is the number of efolds since the 
cosmologically relevant modes exit the horizon till the end of inflation 
and $ \chi $ is a dimensionless, slowly varying field 
$$
\chi = \frac{\varphi}{\sqrt{N} \;  M_{Pl}}  \; .
$$
The slow-roll expansion becomes in this way a systematic $ 1/N $ 
expansion. The couplings in the inflaton Lagrangian become naturally small
due to suppression factors arising from eq.(\ref{VIn}) as
the ratio of the two relevant energy scales here: $ M $ and $ M_{Pl} $.
The spectral index, the ratio of tensor/scalar fluctuations, the running 
index and the amplitude of the scalar adiabatic fluctuations are 
naturally 
\be \label{predi}
n_s - 1 = {\cal O}\left(\frac1{N}\right) \quad , \quad 
r = {\cal O}\left(\frac1{N}\right) \quad , \quad
\frac{d n_s}{d \ln k} =  {\cal O}\left(\frac1{N^2}\right) \quad , \quad
|{\Delta}_{k\;ad}^{\mathcal{R}}|  \sim N \; 
\left(\frac{M}{M_{Pl}}\right)^2 \; ,
\ee
for {\bf all} inflaton potentials in the class of eq.(\ref{VIn}). Hence, 
the energy scale of inflation $ M $ is completely determined by the 
amplitude of the scalar adiabatic fluctuations 
$ |{\Delta}_{k\;ad}^{\mathcal{R}}| $ and using the WMAP5 results for it,
we find $ M \sim 10^{16}$ GeV.
The running index results $ d n_s/d \ln k \sim 1/N^2 \sim  10^{-4} $ and 
therefore can be neglected. Namely, within the class of models 
eq.(\ref{VIn}) one does not need to measure the ratio $ r $ in order to 
learn about the scale of inflation. Moreover, we were able to provide 
{\bf lower} bounds for $ r $ and {\bf predict its value} in the effective 
theory of inflation using the CMB+LSS data and Monte Carlo
Markov  Chains (MCMC) simulations \cite{nos,mcmc}.}
\item{Besides its simplicity, the fourth order
potential (minimal single field model in the Ginsburg-Landau spirit)
is rich enough to describe the physics of inflation and
accurately reproduce the WMAP data \cite{mcmc,nos}. It is well motivated
within the Ginsburg-Landau approach as an effective field theory
description (see ref.\cite{gl,quir}). We provided a complete analytic
study complemented by a statistical analysis in \cite{mcmc,nos}. The MCMC 
analysis of the available CMB+LSS data with the 
Ginsburg-Landau effective field theory of inflation showed \cite{mcmc,nos}:

(i) The data strongly indicate a spontaneous {\bf breaking}
of the $ \varphi \to - \varphi $ symmetry of the
inflaton potentials. Namely, $ w''(0) < 0 $ is strongly favoured.
Cubic terms do not improve the fit.
(ii) Fourth order double-well potentials naturally satisfies this requirement and
provide an excellent fit to the data. (iii) The above results and 
further physical analysis lead us to conclude that {\bf new inflation}
gives the best description of the data. (iv) We find a {\bf  lower bound}
for $ r $ within fourth order double-well potentials: $ r > 0.023 \;
(95\% \; {\rm CL}) $ and $ r > 0.046 \;  (68\% \;   {\rm CL}) $, 
see fig. \ref{banana}. (v) The preferred new inflation potential is a double
well, even function of the field with a moderate quartic coupling 
$ y \sim 1 $,
\be\label{bini}
w(\chi) = \frac{y}{32} \left(\chi^2 - \frac8{y}\right)^{\! 2} 
= -\frac12 \; \chi^2 + \frac{y}{32} \; \chi^4 + \frac2{y}\quad   .
\ee
[see eq.(\ref{VIn})]. This new inflation model yields as most probable 
values: $ n_s \simeq 0.964 ,\; r\simeq 0.051 $, see fig. \ref{banana}. 
This value for $ r $ is within reach of forthcoming CMB observations 
\cite{mcmc,nos}. For the best fit value $ y \simeq 1.26 $,
the inflaton field exits the horizon in the negative
concavity region $ w''(\chi) < 0 $ intrinsic to new inflation.
We find for the best fit, $ M = 0.543 \times 10^{16}$ GeV for the scale of 
inflation and $ m = 1.21 \times 10^{13}$ GeV for the inflaton mass. 
We derived explicit formulae and study in detail the spectral index
$ n_s $ of the adiabatic fluctuations, the ratio $ r $ of tensor to
scalar fluctuations and the running  index $ d n_s/d \ln k $
\cite{mcmc,nos}. We use these analytic formulas as hard constraints on $
n_s $ and $ r $ in the MCMC analysis. Our analysis differs in this
{\bf crucial} aspect from previous MCMC studies in the literature
involving the CMB data. }

\begin{figure}[h]
\includegraphics[height=7.cm,width=12.cm]{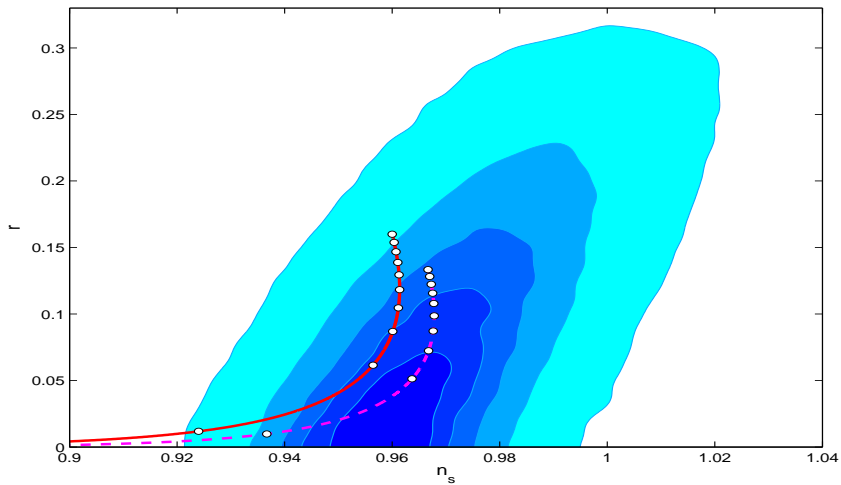}
\caption{$(n_s,\,r)$ from the fourth order double-well inflaton potential 
eq.(\ref{bini}) compared to CMB+LSS data. The color--filled areas correspond to $ 12\%, \; 
27\%, \;  45\%, \;  68 \% $ and $ 95 \% $ confidence levels according to the WMAP, SN
and Sloan data. The color of the areas goes from the darker to the lighter for increasing CL.
The solid red curve is for $ N=50 $ and the dashed 
magenta curve for $ N=60 $. The quartic coupling $ y $ increases 
monotonically starting from the uppermost dots, corresponding to the
free-field, purely quadratic inflaton potential $ y = 0 $ till
the strong coupling region $ y \gg 1 $ in the lower left part of the curve.
We see that very small values of $ r $ {\bf are excluded} since they 
correspond to $ n_s < 0.92 $ outside the $ 95 \% $ confidence level 
contour. That is, we obtain a {\bf lower bound} for $ r : \;
r > 0.027 $ at 95\% C. L.}
\label{banana}
\end{figure}

\item{The effects of arbitrary {\bf higher order} terms in the inflaton 
potential on the CMB observables: spectral index $ n_s $ and ratio $ r $ 
were systematically analyzed in ref. \cite{high}.
The theoretical values in the $ (n_s,r) $ plane for all double well 
inflaton potentials in the Ginsburg-Landau approach turn to be inside an
{\bf universal} banana-shaped region $ \cal B $ displayed in fig. \ref{banana2}. 
The upper border of the 
banana-shaped region $ \cal B $ is given by the fourth order double--well 
potential eq.(\ref{bini}) and provides an upper bound for the ratio $ r $. The lower 
border of $ \cal B $ is defined by the quadratic plus an infinite barrier 
inflaton potential and provides a {\bf lower bound} for the ratio $ r $
within the Ginsburg-Landau class of potentials \cite{high}. 
For example, the current best value of the spectral index $ n_s = 0.964 $, 
implies  $ r $ is in the interval: $ 0.021 < r < 0.053$. Interestingly 
enough, this range is within reach of forthcoming CMB observations.} 

\item{The dynamics of inflation is usually described by the classical 
evolution of a scalar field (the inflaton). The use of classical dynamics
 is justified by the enormous stretching of
physical lengths during inflation. When the physical wavelength of
the fluctuations become larger than the Hubble radius, these
fluctuations effectively become classical. This is probably the only
case where the time evolution itself leads to the classicalization
of fluctuations and microscopic scales near the Planck scale 
$ 10^{-32} \, {\rm cm} \lesssim \lambda = 2 \, \pi / k \lesssim
10^{-28} $ cm become macroscopic today in the range $ 1 \, {\rm
Mpc} \lesssim \lambda_{today} \lesssim 10^4 $ Mpc. This happens thanks to 
a redshift by $ \sim 10^{56} $ since the beginning of inflation for a total
number of inflation efolds $ N_{tot} \sim 64 $.

The effective theory of inflation is generically valid as long as the 
energy density is $ \ll M_{Pl}^4 $ (sec. \ref{iifd}). 
This is true thanks to eq.(\ref{VIn}) even when the inflaton field 
$ \varphi $ takes values equal to many times $  M_{Pl} $ \cite{nos,1sN}.

We computed relevant quantum loop corrections to inflationary dynamics
in ref. \cite{nos,effpot,quant}. Novel phenomena emerge at the quantum
level as a consequence of the lack of kinematic thresholds, among
them the phenomenon of inflaton decay into its own quanta. A
thorough study of the effect of quantum fluctuations reveals that
these loop corrections are suppressed by powers of $
\left(H/M_{Pl}\right)^2 $ where $ H $ is the Hubble parameter during 
inflation \cite{nos,effpot,quant}.
The amplitude of temperature fluctuations constrains the scale of
inflation with the result that $ \left(H/M_{Pl}\right)^2 \sim
10^{-9} $. Therefore, quantum loop corrections are very small
and controlled by the ratio $ \left(H/M_{Pl}\right)^2 $, a
conclusion that  validates the reliability of the classical
approximation and the effective field theory approach to
inflationary dynamics. The quantum corrections to the power spectrum
are computed and expressed in terms of $ n_s, \; r $ and $ dn_s/d \ln k $.
Trace anomalies dominate the quantum corrections to the primordial
power spectrum (see \cite{nos,effpot,quant}). }

\begin{figure}[h]
\includegraphics[height=7.cm,width=12.cm]{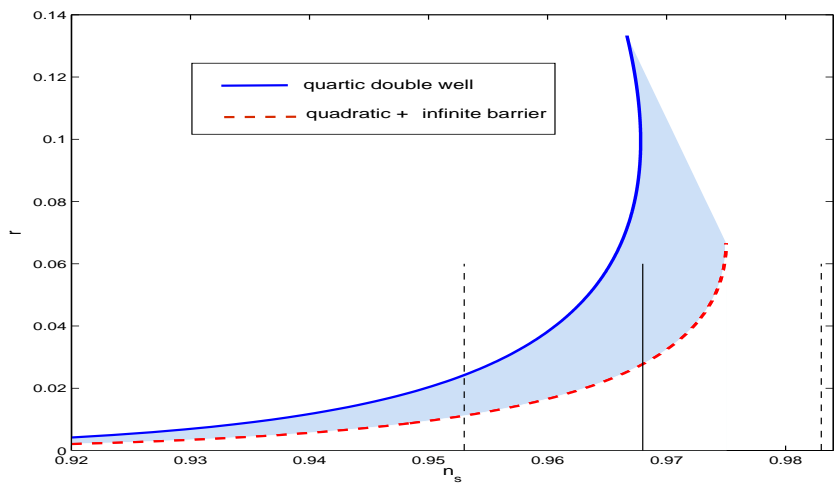}
\caption{We plot here the borders of the universal 
banana region $ \cal B $ in the $ (n_s, r) $-plane 
setting $ N = 60 $. The curves are computed with the quadratic 
plus quartic potential eq.(\ref{bini}) and with  the quadratic plus an 
infinite barrier inflaton potential \cite{high}.}
\label{banana2}
\end{figure}

\item{Scalar (curvature) and tensor (gravitational wave) perturbations 
originate in quantum fluctuations during inflation. These are usually 
studied within the slow-roll approximation and with Bunch-Davies
initial conditions. We investigated the physical effects on the power 
spectrum of generic initial conditions with particular attention to 
back-reaction effects in refs. \cite{quadru1,quadru2}. We introduced a
\emph{transfer function } $ D(k) $ which encodes the effect of
generic initial conditions on the power spectra. The constraint from
renormalizability and small back reaction entails that $ D(k)
\lesssim \mu^2/k^2 $ for large $ k $ where $ \mu $ characterizes the 
asymptotic decay of the occupation number. This implies that observable
effects from initial conditions are more prominent in the \emph{low}
CMB multipoles.  The effects on high $l$-multipoles are
suppressed by a factor $ \sim 1/l^2 $ due to the large $ k $ fall off  of 
$ D(k) $. Hence, a change from the Bunch-Davies initial conditions for the 
fluctuations can naturally account for the low observed value
of the CMB quadrupole \cite{quadru1,quadru2,quamc,singu}.}

\item{The inflaton evolution generically starts by a 
fast-roll stage where the kinetic and potential
energy of the inflaton are of the same order \cite{quadru1,quadru2,nos}.
The universe expansion is non-inflationary to start (decelerated)
and hence no fluctuations leave the horizon then. This leads to a 
suppression of the quadrupole in curvature and tensor perturbations 
\cite{quadru1,quadru2,quamc,nos}. The fast-roll stage becomes then
accelerated (inflationary) and is followed by a slow-roll regime 
where the kinetic energy is much smaller than the potential energy. 
The slow-roll regime of inflation is 
an attractor of the dynamics \cite{bgzk} during which
the Universe is dominated by vacuum energy. Inflation ends
when again the kinetic energy of the inflaton becomes large as the
field is rolling near the minimum of the potential. Eventually,
the energy stored in the homogeneous inflaton is transferred
explosively into the production of particles via spinodal or
parametric instabilities \cite{chalo,prehea,baacke,ramsey}. 
More precisely, non-linear phenomena eventually 
{\bf shut-off} the instabilities and {\bf stop} inflation 
\cite{chalo,chalo2,tsu}.
All these processes lead to the transition to a radiation dominated era.
This is the standard picture of the transition from
inflation to standard hot big bang cosmology.}

\item{Within the context of the effective 
field theory and for generic initial conditions on the inflaton field, it
is shown that a quadrupole suppression consistent with observations
is a natural consequence of the fast-roll stage 
\cite{quadru1,quadru2,quamc}. The fast-roll
stage dynamically modifies the initial power spectrum of perturbations
by a transfer function $ D(k) $. We performed MCMC analysis of the WMAP 
and SDSS data combined with the most recent supernovae compilation 
\cite{SN} including the fast-roll stage. The quadrupole mode 
$ k_Q = 0.238 \; {\rm Gpc}^{-1} $ exits the horizon about 0.2
efolds before the end of fast-roll. This fixes
the redshift since the beginning of inflation till today to 
$ z_{init} = 0.915 \times 10^{56} $. From that we find the {\bf 
total number of efolds} $ N_{tot} $ during inflation to be (see 
ref. \cite{nos,singu}) $ N_{tot} \simeq 64 $ respecting the lower bound 
for $ N_{tot} $ that solves the horizon problem  (see  \cite{nos}).
That is, the MCMC analysis of the CMB+LSS data including the early 
fast-roll explanation of the CMB quadrupole suppression {\bf imposes} 
$ N_{tot} \simeq 64 $. Including the fast-roll stage improves the fits 
to the TT, the TE and the EE modes, well reproducing the quadrupole 
supression. }
\end{itemize}

We formulate inflation as an effective field theory within the 
Ginsburg-Landau spirit \cite{1sN,gl,nos}. The theory of the second order 
phase transitions, the Ginsburg-Landau theory of superconductivity, 
the current-current Fermi theory of weak interactions, the sigma model of 
pions, nucleons (as skyrmions) and photons are all successful
effective field theories. Our work shows how powerful is
the effective theory of inflation {\bf to predict observable quantities} 
that can be or will be soon contrasted with experiments. 
There are {\bf two kind} of predictions in the effective theory of 
inflation: first, predictions on the order of magnitude of the CMB
observables valid for all inflaton potentials in the class of 
eq.(\ref{VIn}) as given by eqs.(\ref{predi}); second, precise quantitative 
predictions as those presented in refs. \cite{nos,mcmc,quamc,high,singu}.

The Ginsburg-Landau realization of the inflationary potential 
fits the amplitude of the CMB anisotropy remarkably well
and reveals that the
Hubble parameter, the inflaton mass and non-linear couplings are
see-saw-like, namely powers of the ratio $ (M/M_{Pl})^2 \sim 10^{-9} $ 
multiplied by further powers of $ 1/N $ \cite{nos,1sN}. 
Therefore, the smallness of the 
couplings is not a result of fine tuning but a {\bf natural} consequence of
the form of the potential, of the validity of the effective field theory
description and slow-roll (sec. \ref{potuniv}). 
The quantum expansion in loops is
therefore a double expansion on $ \left(H/M_{Pl}\right)^2 $ and $
1/N $. Notice that graviton corrections are also  of
order $ \left(H/M_{Pl}\right)^2 $ because the amplitude of tensor
modes is of order $H/M_{Pl}$. 

The infrared (superhorizon) modes in the quantum loops produce large 
contributions of the order $ \sim N $. However, as shown in
\cite{nos,effpot,quant} these large infrared contributions get 
multiplied by 
slow-roll factors of order  $ \sim 1/N $. As a result, the 
superhorizon contributions to physical magnitudes turn to be of
order $ N^0 $ \cite{nos,effpot,quant} times  factors of the order of
$ \left(H/M_{Pl}\right)^2  \sim 10^{-9} $.

We note that the effective theory of inflation describes an evolution
spanning about 26 orders of magnitude in length scales from the beginning 
till the end of the inflationary era. This is the largest scale change 
described by a field theory so far.

It must be stressed that the energy scale of inflation,
$ M \sim 10^{16} $ GeV is the energy scale of at least two
other important physical situations: (a) the scale of Grand Unification
of strong and electroweak interactions and (b) the large energy scale 
in the see-saw formula for neutrino masses [see eq.(\ref{neum})].
This coincidence suggests a physical link between the three areas.

\medskip

Many deep problems remain to be solved in the early universe. One of 
them is the reheating problem. Namely, how the universe thermalizes 
after inflation and at which temperature. Baryogenesis provides a lower 
bound on the reheating temperature \cite{kt}. The mechanisms of 
thermalization uncovered in refs. \cite{fi4} can 
provide a starting point to understand the reheating.
The units used here are such that $ \hbar = c = 1 $.

\section{The Standard Model of the Universe}\label{uno}

The history of the Universe is a history of expansion and cooling down.

On large scales the Universe is homogeneous and isotropic and its
geometry is described by the spatially flat Friedmann-Robertson-Walker 
(FRW) metric
\be\label{FRW}
 ds^2= dt^2-a^2(t) \; d\vec{x}^2
\ee
Overwhelming observational evidence indicates that the geometry 
of the Universe is spatially flat. Namely, in case the space is curved, its
curvature radius is larger than the horizon and therefore inobservable.

Notice that this cosmological expansion has no 
center: it happens everywhere at all spatial points $ \vx $ and it is 
identical everywhere. The scale factor grows monotonically with time.

Physical scales are stretched by the scale factor $ a(t) $ with respect to 
the time independent comoving scales
\begin{equation}\label{scale}
l_{phys}(t)= a(t) \; l_{com} \; .
\end{equation}
The redshift $ z $ at time $ t $ is defined as
\be\label{z}
z + 1 \equiv \frac1{a(t)}
\ee
where the scale factor today is choosen to be unit 
$ a(0) \equiv 1 $. The farther back in time, the larger is the redshift 
and the smaller is $ a(t) $.

The temperature decreases as the universe expands as
\be \label{Ta} 
T(t) = \frac{T_0}{a(t)}  \; .
\ee 
Eq.(\ref{Ta}) applies to all particles in thermal equilibrium
as well as to massless decoupled particles (radiation).
Since the temperature decreased with time, the Universe underwent a 
succession of phase transitions towards the less symmetric phases 
\cite{tranfas}.

\medskip

The combination of data from CMB and LSS, and numerical
simulations lead to the $\Lambda$CDM or \emph{concordance model}
which has now become the standard cosmology. This impressive
convergence of observational data and theoretical and numerical results
describes a Universe that is composed of a cosmological constant, dark
matter, baryonic (atoms) matter and radiation. This model
provides the {\bf only consistent} explanation of  the
broad set of precise and independent astronomical observations over
a wide range of scales available today. Namely:

\begin{itemize}

\item{WMAP data and previous CMB data.}

\item{Light Elements Abundances.}

\item{Large Scale Structures (LSS) Observations. Baryon acoustic 
oscillations (BAO).}

\item{Acceleration of the Universe expansion:
Supernova Luminosity/Distance (SN) and Radio Galaxies.}

\item{Gravitational Lensing Observations.}

\item{Lyman $ \alpha $ Forest Observations.}

\item{Hubble Constant ($H_0$) Measurements.}

\item{Properties of Clusters of Galaxies.}

\item{Measurements of the Age of the Universe.}

\end{itemize}

In the homogeneous and isotropic FRW universe described by eq.(\ref{FRW}), the
matter distribution must be homogeneous and isotropic, with an
energy momentum tensor having in spatial average the isotropic fluid form 
\be\label{fluido}
\langle T^{\mu}_{\nu} \rangle={\rm diag}[\rho, -p,-p,-p ] \; , 
\ee
where $ \rho, \; p $ are the energy density and pressure,
respectively. In such space-time geometry the Einstein equations of general
relativity reduce to the Friedmann equation, which determines
the evolution of the scale factor from the energy density
\be\label{fri}
\left[\frac{\dot{a}(t)}{a(t)} \right]^2 = H^2(t) = \frac{
\rho}{3 M^2_{Pl}} \,.
\end{equation}
where $ M_{Pl}= 1/\sqrt{8\pi G} = 2.43534 \times 10^{18}$ GeV
$ = 0.434\times 10^{-5}$ g.
The spatially flat Universe has today the critical density
\be\label{rhocrit}
\rho_c = 3 \; M^2_{Pl} \; H^2_0= 1.878 \; h^2 \;
10^{-29}\textrm{g}/\textrm{cm}^3 = 1.0537 \; 10^{-5} \;  
h^2 \; \textrm{GeV}/\textrm{cm}^3 = (2.518 \; {\rm meV})^4 \; .
\ee
where $ H_0 = 100 \; h \; \textrm{km/sec}/\textrm{Mpc} $ is the Hubble 
constant today, $ h = 0.705 \pm 0.013 $ \cite{pdg,WMAP5} and then
$ H_0 = 1.5028 \; 10^{-33} $ eV, 1 meV $ = 10^{-3} $ eV.
Notice that eq.(\ref{fri}) implies that $ a(t) $ is a monotonic
function of time.

The energy momentum tensor conservation reduces to the single
conservation equation,
\begin{equation}\label{conener}
\dot{\rho}+3 \; H(t) \; \left( \rho+p\right)  =0
\end{equation}
\noindent The two equations (\ref{fri}) and (\ref{conener}) can
be combined to yield the acceleration of the scale factor,
\begin{equation}\label{accel}
\frac{\ddot{a}}{a}= -\frac1{6 \; M^2_{Pl}}(\rho + 3 \; p) \; .
\end{equation}
In order to provide a close set of equations we must append an
equation of state $ p=p(\rho) $ which is typically written in the form
\begin{equation}\label{eqnofstate}
p=w(\rho) \; \rho
\end{equation}

\begin{table}
\begin{tabular}{|c|c|c|c|c|}
\hline  $\rho_c$  &  $ (2.36 \;  {\rm meV})^4 $ & & $h$   &  
$ 0.705 \pm 0.013 $ \\
\hline $ H_0 $ & $ h /[3 \; {\rm Gpc}] =h /[9.77813 \; {\rm Gyr}]$ & & 
$\Omega_{\Lambda}$ & $0.726$ \\ 
\hline  $ M_{Pl} $  & $2.43534\times 10^{18}$ GeV & &  $\Omega_M $ & $0.274$ \\ 
\hline  $ M $  & $0.543 \times 10^{16}$ GeV & &  $\Omega_r $ & 
$8.49 \; 10^{-5}$ \\ \hline  $ m $  & $1.21 \times 10^{13}$ GeV& &$ n_s $ & 
$ 0.960 \pm 0.014 $\\ \hline
\end{tabular} 
\caption{Selected Cosmological Parameters \cite{pdg,WMAP5}. $ m $ and $ M $
are given by eq.(\ref{myM}).}
\end{table}

The following are  important cosmological solutions:
\bea\label{fa}
&&  {\rm Cosmological ~ Constant } \; \Rightarrow
p=-\rho : {\bf\Lambda D}{\rm ~de~Sitter~ expansion}\Rightarrow \rho = {\rm
constant}~;~ a(t) = a(0) \; e^{Ht} ~;~H= 
\sqrt{\rho/[3 \; M^2_{Pl}]} \cr \cr
&& {\rm Radiation}\; \Rightarrow p/\rho=1/3:
{\rm {\bf RD} \; (Radiation ~ domination)}
\Rightarrow \rho(t)= \rho(t_r) \;  a^{-4}(t)~;~
a(t) = a(t_r) \; \sqrt{t/t_r} \label{RD} \\
&& {\rm Non-relativistic ~ (cold) ~ Matter} \; \Rightarrow p/\rho=0 :
{\rm {\bf MD} \;  (Matter ~ domination)}  \Rightarrow \rho(t) =
\rho(t_{eq}) \; a^{-3}(t)~;~ a(t) = a(t_{eq}) \; 
(t/t_{eq})^\frac23 \nonumber 
\eea 
where  $ t_r $ and $ t_{eq} $ are the values of cosmic time at which the 
Universe becomes radiation or matter dominated, respectively.

Notice from eq.(\ref{accel}) that
accelerated expansion ($ {\ddot a}(t) > 0 $) takes place if $ p/\rho < -1/3 $. 

\medskip

The universe started by a very short accelerated inflationary stage 
dominated by the vacuum energy, lasting $ \sim 10^{-36}$ sec
ending by redshift $ z \sim 10^{29} $ and approximately described by the 
de Sitter metric. This inflationary stage was followed by decelerated 
expansion, first by the radiation dominated era and then by the matter 
dominated era. Finally, the universe entered again an accelerated phase 
dominated by the dark energy, described by a cosmological constant in the 
Standard Model of the Universe, at $ z \simeq 0.5 $.

\medskip

Particle physics at energy scales below $ \sim 200 $ GeV is on solid 
experimental footing in the framework of the standard model of strong and 
electroweak interactions.

Current theoretical ideas supported by the renormalization group
running of the couplings in the standard model of particle physics
and its supersymmetric extensions show that the strong, weak and
electromagnetic interactions are unified in a grand unified theory
(GUT) at the scale $ M_{GUT} \sim 10^{16}$ GeV. 
Furthermore, the characteristic scale at which gravity calls for a quantum
description is the Planck scale $  M_{Pl} = 1/\sqrt{8\pi G} = 2.43534 \; 10^{18}$ GeV
$ \gg M_{GUT} $.

The connection between the standard model of particle physics and
early Universe cosmology is through the semiclassical 
Einstein equations that
couple the space-time geometry to the matter-energy content. As
argued above, gravity can be studied semi-classically at energy
scales well below the Planck scale. The standard model of particle
physics is a {\em quantum field theory}, thus the space-time is
classical  but with sources that are quantum fields. Semiclassical
gravity is defined by the Einstein's equations with the
expectation value of the quantum energy-momentum tensor $ \hat{T}^{\mu
\nu} $ as the source
\begin{equation}\label{einstein}
G^{\mu \nu} = R^{\mu \nu} -\frac12 \; g^{\mu \nu} R = \frac{
\langle \hat{T}^{\mu \nu} \rangle}{M^2_{Pl}} \; .
\end{equation}
The expectation value of  $ \hat{T}^{\mu \nu} $ is taken in a
given quantum state (or density matrix) compatible with
homogeneity and isotropy which must  be translational and
rotational invariant. Such state yields an expectation value for
the energy momentum tensor with the fluid form eq.(\ref{fluido}),
and the Einstein equations (\ref{einstein}) reduce to the Friedmann
equation  (\ref{fri}).

All of the ingredients 
are now in place to understand the evolution of the early Universe.
Einstein's equations determine the evolution of the scale factor, particle physics
provides the energy momentum tensor and statistical mechanics
provides the fundamental framework to describe the thermodynamics
from the microscopic quantum field theory of the strong, 
electroweak interactions and beyond.

The sources for Einstein equations are dark energy, dark and ordinary 
matter and radiation. The standard model of particle physics describes 
ordinary matter and radiation. 

Dark energy accounts {\bf today} for $ 72 \pm 1.5\% $ of the energy of the
Universe \cite{WMAP5}. The current observed value is $ \rho_{\Lambda} =  
\Omega_{\Lambda} \; \rho_c = (2.36 \;  {\rm meV})^4 $ from eq.(\ref{rhocrit})
\cite{pdg,WMAP5}. 
The equation of state is $ p_{\Lambda} = - \rho_{\Lambda} $ within 
observational errors corresponding to a cosmological constant. 

The nature of the dark energy (today) is not 
yet understood. A plausible explanation of the dark energy
may be the quantum zero-point energy of a light matter field in the cosmological
space-time. This has the equation of state of a cosmological constant. 
Notice that the renormalized value of the zero point energy in the cosmological
space-time is finite and may be naturally of the order of the (mass)$^4$ of the 
light field involved.

Matter accounts today for $ 28 \pm 1.5\% $ of the energy of the Universe 
\cite{WMAP5}. $ 84 \% $ of the matter is {\bf dark matter}.
Therefore,  dark matter is an essential constituent of the universe.
The nature of dark matter is still unknown but is certainly
beyond the Standard Model of strong and electroweak particle interactions 
\cite{libros,tranfas}. It is probably formed by particles in the keV mass
scale as discussed in sec. \ref{secdm} \cite{oscnos,mnras}.

\begin{figure}[h]
\begin{center}
\begin{turn}{-90}
\psfrag{Omega_{Lambda} vs. log(1+z)}{$ \frac{\rho_{\Lambda}}{\rho}$ vs. 
$ \log(1+z) $}
\psfrag{Omega_{mat} vs. log(1+z)}{$ \frac{\rho_{Mat}}{\rho} $ vs. 
$ \log(1+z) $}
\psfrag{Omega_{rad} vs. log(1+z)}{$ \frac{\rho_{rad}}{\rho} $ vs. 
$ \log(1+z) $}
\includegraphics[height=13.cm,width=7.cm]{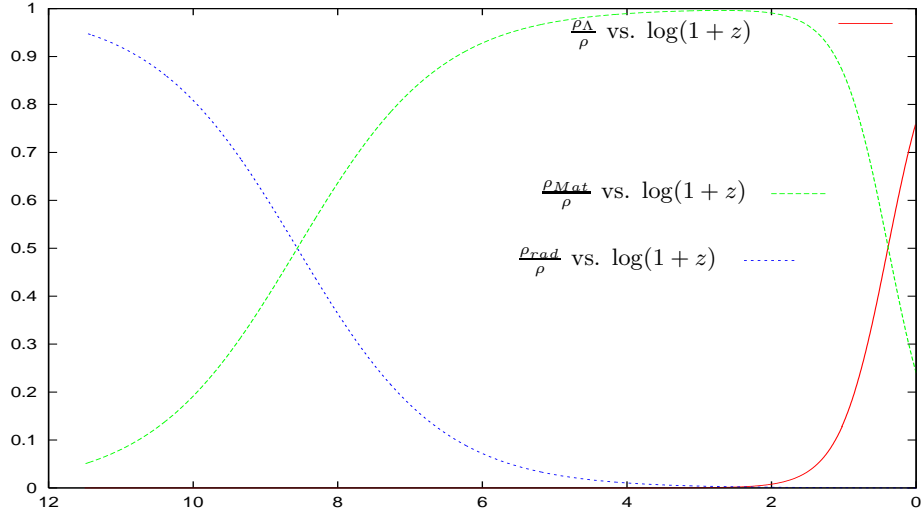}
\end{turn}
\caption{ $ \frac{\rho_{\Lambda}}{\rho}, \; \frac{\rho_{Matter}}{\rho} $ and 
$ \frac{\rho_{radiation}}{\rho} $ vs.  $ \log(1+z) $.} 
\label{ome}
\end{center}
\end{figure}

Main events in the universe after inflation are (see fig. \ref{ome}): 

\begin{itemize}
\item{Begining of the {\bf RD} era and end of inflation:  $ z \sim 10^{29} 
\; , \quad T_{reh} \sim 10^{16} $ GeV, $ t  \sim 10^{-36}$ sec.}

\item{Electro-Weak phase transition: $ z \sim 10^{15} \; , \quad
T_{\rm EW} \sim 100$ GeV, $ t\sim 10^{-11} $ sec.}

\item{QCD phase transition (confinement): $ z \sim 10^{12} \; , 
\quad T_{\rm QCD} \sim 170$ MeV, $ t \sim 10^{-5} $ sec.}

\item{Big bang nucleosynthesis (BBN):  $ z \sim  10^9 \; , \quad \ln(1+z)\sim 
21  \; , \quad  T \simeq 0.1$ MeV, $ t \sim 20 $ sec.}

\item{Radiation-Matter equality: $ z \simeq 3200  \; , \; \ln(1+z) 
\simeq 8 \; , \;  T \simeq 0.7$ eV,  $ t \sim 57000 $ yr.}

\item{CMB last scattering: $ z \simeq 1100 \; , \; \ln(1+z) \simeq 7  
\; , \;  T \simeq 0.25$ eV,  $ t \sim 370000 $ yr.}

\item{Matter-Dark Energy equality: $ z  \simeq 0.47  \; , \; \ln(1+z) 
\simeq 0.38 \; , \;  T \simeq 0.345$ meV $ t \sim 8.9 $ Gyr.}

\item{Today: $ z = 0 \; , \; \ln(1+z) = 0 \; , \;  T = 2.725$K $=0.2348 $ 
meV,  $ t \equiv t_0 = 13.72 $ Gyr.}
\end{itemize}

In fig. \ref{ome} we plot $ \rho_{\Lambda}/\rho, \; \rho_{Matter}/\rho $ 
and $ \rho_{radiation}/\rho $ as functions of $ \log(1+z) $ where 
$ \rho_{\Lambda} =\Lambda,
\; \rho_{Matter} = \Omega_M/a^3 $ and $ \rho_{radiation} = \Omega_r/a^4 $.
Notice that $ \rho_{\Lambda} + \rho_{Matter} +  \rho_{radiation} = \rho $.

\medskip

In summary, the Friedmann equation (\ref{fri}) can be written as
\be\label{frgen}
H^2(t) = H^2_0 \left[\Omega_{\Lambda} + \frac{\Omega_M}{a^3} 
+ \frac{\Omega_r}{a^4}\right] \; .
\ee
The temperature of the universe in the post-inflation radiation dominated era 
(reheating temperature $ T_r $) is bounded from below in order to explain the 
baryon asymmetry and the Big Bang Nucleosynthesis (BBN). This amounts
to a further constraint on the inflationary model. The BBN constraint
is the milder. If  the observed baryon asymmetry is
produced at the electroweak scale, the  constraint on the  reheating
temperature is $ \gtrsim 100 $ GeV, however the origin of the baryon
asymmetry may be at the GUT scale in which case the reheating temperature 
should be $  T_r > 10^9 $GeV \cite{kt}.

\section{Inflation and Inflaton field dynamics}\label{iifd}

A simple implementation of the inflationary scenario is based on a single
scalar field, the \emph{inflaton} with a Lagrangian density 
\be
\mathcal{L} = a^3(t)\left[\frac{\dot{\varphi}^2}2 -
\frac{(\nabla\varphi)^2}{2a^2(t)}-V(\varphi) \right] \; ,
\ee 
where $ V(\varphi) $ is the inflaton potential. Since the universe
expands exponentially fast during inflation, gradient terms 
are  exponentially suppressed and can be neglected.
At the same time, the exponential stretching of spatial lengths
classicalize the physics and permits a classical treatment.
One can therefore consider an homogeneous and classical inflaton field 
$ \varphi(t) $ which obeys the evolution equation
\be\label{eqno} 
{\ddot \varphi} + 3 \, H(t) \; {\dot \varphi} + V'(\varphi) = 0 \; . 
\ee 
in the isotropic and homogeneous FRW metric eq.(\ref{FRW}) which is sourced by the 
inflaton according to the Friedmann equation eq.(\ref{fri}). Eq.(\ref{fri}) 
takes here the form
\be\label{frinf}
H^2(t) = \frac1{3 M^2_{Pl}} \left[\frac12 \; \dot \varphi^2 + V(\varphi)\right] \; .
\ee
The energy density and the pressure for a spatially homogeneous inflaton 
are given by
\be\label{enerpres} 
\rho = \frac{\dot{\varphi}^2}2+ V(\varphi)
\quad , \quad p  =\frac{\dot{\varphi}^2}2-V(\varphi) \; . 
\ee
The time derivative of the Hubble parameter takes the form
\be
{\dot H}(t) = -\frac{\dot{\varphi}^2}{2 \;  M^2_{Pl}}
\ee
where we used eqs.(\ref{eqno}) and (\ref{frinf}).
This shows that $  H (t) $ {\bf decreases monotonically} with time.

The inflaton fields starts with some chosen values of $ \varphi $ 
and  $ \dot \varphi $ and evolves together with the scale factor according to
eqs.(\ref{eqno}) and (\ref{frinf}). The inflaton clearly rolls down the slope
of the potential going towards a local minimum of $ V(\varphi) $.
The basic constraint on the inflationary potential is
\be\label{vmin}
V(\varphi_{min})= V'(\varphi_{min})= 0 \; .
\ee
That is, the inflaton potential {\bf must vanish} at its minimum $ \varphi_{min} $
in order to have a finite number of efolds.
The inflaton evolves from its initial value 
towards the minimum $ \varphi_{min} $. If $ V(\varphi_{min}) > 0 $, we see from 
eq.(\ref{frinf}) that inflation will be eternal. That is, a de Sitter phase will 
continue forever with the inflaton at the constant value $ \varphi_{min} $.

There are two main classes of inflaton potentials leading to two main  
classes of inflation.

\noindent
(a) In {\bf small field inflation} the minimum of the potential is at a 
non-zero value $ \varphi_{min} \neq 0 $ and the inflaton field starts near 
(or at) $ \varphi = 0 $ evolving towards $ \varphi =\varphi_{min} $.
These are discrete symmetry ($ \varphi \to - \varphi $) breaking 
potentials \cite{infnue}, 
\be\label{nueva}
V(\varphi) = \frac{\lambda}4 \left( \varphi^2- 
\frac{m^2}{\lambda} \right)^2 =
-\frac{m^2}2 \; \varphi^2 +\frac{\lambda}4  \; \varphi^4 + 
\frac{m^4}{4 \; \lambda} \quad , \quad {\rm new \; inflation} \; .
\ee
(b) In {\bf large field inflation} the minimum of the potential is at 
$ \varphi_{min} = 0 $ and the inflaton field starts  at $ \varphi \gg M $ 
evolving towards  $ \varphi = 0 $.
These are unbroken symmetry potentials \cite{infcao}, 
\be\label{caotica}
V(\varphi)= +\frac{m^2}2  \; \varphi^2 +\frac{\lambda}4  \; \varphi^4 
\quad , \quad {\rm chaotic \; inflation} \; .
\ee
For historical reasons large field inflation is often called {\bf  chaotic inflation}
and small field inflation {\bf new inflation}.

\medskip

Inflation should last at least $ N_{tot} \gtrsim 64 $ efolds in order to solve the entropy, 
horizon and flatness problems (see, for example, ref. \cite{nos}). Inflation can produce 
such large number of efolds provided it lasts enough time. This can be achieved if the 
inflaton evolves slowly (slow-roll), namely $ \dot{\varphi}^2 \ll V(\varphi) $. 
This implies from eq.(\ref{enerpres}) that
$$
\rho = -p \simeq  V(\varphi) \simeq  {\rm constant},
$$
as the equation of state leading to a de Sitter universe. 
Eq.(\ref{frinf}) yields as scale factor 
\be\label{groso}
a(t) \simeq e^{H \; t} \quad,\quad H \simeq \sqrt{\frac{V(\varphi)}{3 \; M^2_{Pl}}} 
\ee
[see eq.(\ref{fa})]. However, eq.(\ref{groso}) is only an
approximation to the slow-roll inflationary dynamics. The scale factor in
the slow-roll approximation is presented in ref. \cite{nos,singu}.

\medskip

While inflationary dynamics is typically studied in terms of a
\emph{classical} homogeneous inflaton field as explained above, such 
classical field must be understood as the expectation value of a 
\emph{quantum field} in an isotropic and homogeneous quantum state. 

In ref.\cite{chalo,chalo2,tsu} the \emph{quantum dynamics} of inflation was
studied for small and large field inflation.
The  initial quantum state was taken to be a gaussian wave
function(al) with vanishing or non-vanishing expectation value of
the field. This state evolves in time with the full inflationary
potential which features an unstable (spinodal) region for $ \varphi^2 <
m^2/(3 \; \lambda) $ where $ V''(\varphi) < 0 $ in the broken symmetric 
case eq.(\ref{nueva}). Just as in the case of Minkowski space time, there 
is a band of spinodally or
parametrically unstable wavevectors, in which the
amplitude of the quantum fluctuations grows exponentially fast
\cite{tranfas,chalo}. Because of the
cosmological expansion wave vectors are redshifted into the
unstable band and when the wavelength of the unstable modes
becomes larger than the Hubble radius these modes become
\emph{classical} with a large amplitude and a frozen phase. These
long wavelength modes assemble into a classical coherent and
homogeneous condensate, which obeys the equations of motion of the
classical inflaton\cite{chalo,chalo2,tsu}.  This phenomenon of
classicalization and the formation of a homogeneous condensate
takes place during the \emph{first} $ 5-10 $ efolds after the
beginning of inflation. The {\bf non perturbative}
quantum field theory treatment in refs.\cite{chalo,chalo2,tsu} shows that 
this rapid redshift and classicalization justifies the use of an 
homogeneous classical inflaton leading to the following robust
conclusions \cite{chalo,chalo2,tsu}:
\begin{itemize}
\item{ The quantum fluctuations of the inflaton are of two
different kinds: 
\noindent
(a) Large amplitude quantum inflaton fluctuations generated
at the beginning of inflation through spinodal instabilities or parametric
resonance depending on the inflationary scenario chosen. They have
at the beginning of inflation physical 
wavenumbers in the range of \cite{chalo,chalo2,tsu}
\be\label{modosa}
k \lesssim 10 \; m \; ,
\ee
and they become superhorizon a few efolds after the beginning of 
inflation.
The phase of these long-wavelength inflaton fluctuations freeze 
out and their
amplitude grows thereby effectively forming a homogeneous
\emph{classical} inflaton condensate. The study of more general initial
quantum states featuring highly excited distribution of quanta lead
to similar conclusions \cite{tsu}: during the first few efolds of
evolution the rapid redshift produces a classicalization of
long-wavelength  inflaton fluctuations and the emergence of a homogeneous
coherent  inflaton condensate obeying the \emph{classical equations of
motion} in terms of the inflaton potential. 
\noindent
(b) Cosmological scales relevant for the observations \emph{today} 
between $ \sim 1 $ Mpc and the horizon today had 
exited the Hubble radius inside a window of about $ 10 $ e-folds,
from $ \sim 63 $ to $ \sim 53 $ efolds before the end of inflation \cite{kt}.
These correspond to small fluctuations of high physical wavenumbers
at the beginning of inflation in the
range given by \cite{nos}
\be\label{rangok}
3.8 \;  10^{14} \; {\rm GeV} \;  {\beta}^{-1} \; 
e^{N_{tot}-64} < k^{init} <
 3.8 \; 10^{18} \; {\rm GeV} \; {\beta}^{-1} \; e^{N_{tot}-64} \; .
\ee
where $ \beta \equiv \sqrt{10^{-4} \; M_{Pl}/H} \sim 1 $ and $ H $ is 
Hubble by the end of inflation. Since [eq.(\ref{myH})] 
$ m \sim 10^{13} $ GeV,  we see that the large amplitude modes 
eq.(\ref{modosa}) for typical values $ N_{tot} \sim 64 $ and 
$ H \sim 10^{-4} \; M_{Pl} $ correspond to scales {\bf larger}
 than the horizon today.}
\item{During the rest of the inflationary stage the dynamics is described
by this classical homogeneous condensate that obeys the classical 
equations of motion with the inflaton potential. Thus, inflation even if
triggered by an initial quantum state or density matrix of the
quantum field, is effectively described in terms of a classical 
homogeneous scalar condensate. }
\end{itemize}

The body of results emerging from these studies provide a
justification for the description of inflationary dynamics in
terms of a \emph{classical} homogeneous scalar field. The conclusion
is that  after a few initial e-folds during which the unstable
wavevectors are redshifted well beyond the Hubble radius, all
what remains for the ensuing dynamics is a homogeneous classical
condensate, plus small quantum fluctuations corresponding to the wave 
$k$-modes.

These {\bf small} quantum fluctuations include scalar curvature and tensor
gravitational fluctuations. They must be treated
together with the inflaton fluctuations in the unified gauge invariant
approach \cite{hu,nos}.
In the treatment of {\bf large} amplitude quantum inflaton fluctuations,
gravitational fluctuations can be safely neglected \cite{chalo}.

\medskip

Inflation based on a scalar inflaton field should be considered as
an {\bf effective theory}, namely,  not necessarily a fundamental
theory but as a low energy limit of a microscopic fundamental
theory. The inflaton may be a coarse-grained average of
fundamental scalar fields, or a composite (bound state) of fields
with higher spin, just as in superconductivity. Bosonic fields do
not need to be fundamental fields, for example they may emerge as
condensates of fermion-antifermion pairs $ < {\bar \Psi} \Psi> $
in a grand unified theory (GUT) in the cosmological background. In
order to describe the cosmological evolution it is enough to consider
the effective dynamics of such condensates.  

We computed in close form the inflaton potential 
dynamically generated when the inflaton field is a fermion condensate
in the inflationary universe \cite{high}. 
We considered in ref. \cite{high} the inflaton field coupled to Dirac 
fermions $ \Psi $ through the interaction Lagrangian
\be\label{Lf}
{\cal L} = \overline{\Psi}\left[i\,\gamma^\mu \;  \mathcal{D}_\mu  -m_f -
g_Y \; \varphi \right]\Psi \; ,
\ee
Where $ g_Y $ stands for a generic Yukawa coupling and the fermion mass $ m_f $ is
absorbed by a constant shift of the inflaton field. 
The $ \gamma^\mu $ are Dirac $ \gamma $-matrices in curved space-time and 
$ \mathcal{D}_\mu $ stands for the fermionic covariant derivative. 

The effective potential of fermions can be computed in de Sitter inflation
 with the result \cite{quant,nos},
\be\label{Vfermi}
  V_f(\varphi) = V_0 - \frac12 \, m^2 \, \varphi^2 + 
\frac14 \, \lambda \varphi^4 + H^4\,Q\left(g_Y \,\frac{\varphi}{H}\right)\;,
\ee
where,
\be \label{Vfermi2}
    Q(x) = \frac{x^2}{8\,\pi^2}\left\{(1+x^2)\left[\gamma + 
        \mathrm{Re}\,\psi(1+i\,x)\right] - \zeta(3)\,x^2 \right\}
\quad , \quad x \equiv g_Y \,\dfrac{\varphi}{H} \; .
\ee
The constant $ V_0 $ must be such that $ V_f(\varphi) $ fulfills eq.(\ref{vmin}),
$ m^2  > 0 $ and $ \lambda $ are the renormalized mass and 
renormalized coupling constant, $ \psi(x) $ stands for the digamma function, 
$ \gamma $ for the Euler-Mascheroni constant and $ \zeta(x) $ for the Riemann zeta function.
The inflaton potential  $ V_f(\varphi) $ turns to belong to the Ginsburg-Landau class
and provides $ (n_s,r) $ inside the {\bf universal} 
banana-shaped region $ \cal B $ depicted in fig. \ref{banana2} \cite{high}. 

\medskip

The relation between
the low energy effective field theory of inflation and the
microscopic fundamental theory is akin to the relation between the
effective Ginsburg-Landau theory of superconductivity \cite{gl} and the
microscopic BCS theory, or like the relation of the $ O(4) $ sigma
model, an effective low energy theory of pions, photons and nucleons 
(as skyrmions), with quantum chromodynamics (QCD) \cite{quir}. The
guiding principle to construct the effective theory is to include
the appropriate symmetries \cite{quir}. Contrary to the sigma model
where the chiral symmetry strongly constraints the
model \cite{quir}, only general covariance can be imposed on the
inflaton model.

\medskip

In summary, the physics during inflation is characterized by:
\begin{itemize}
\item{Out of equilibrium matter field evolution in a rapidly
expanding space-time
dominated by the vacuum energy. The scale factor is quasi-de Sitter:
 $ a(t) \simeq e^{H \, t}.$}
\item{Extremely high energy density at the scale of 
$\lesssim 10^{16}$ GeV.}
\item{Explosive particle production at the beginning of inflation
due to spinodal or parametric {\bf instabilities} for new and chaotic
inflation, respectively \cite{chalo,chalo2}.}
\item{The enormous redshift as a consequence of a large number of e-folds 
($ \sim 64 $) classicalizes the dynamics: an  {\bf assembly} of
(superhorizon) fluctuations behave as the classical and homogeneous
inflaton field. The inflaton which is a long-wavelength
condensate slowly rolls down the potential hill towards its
minimum \cite{chalo2}.}
\item{Quantum non-linear phenomena eventually {\bf shut-off} the 
instabilities and {\bf stop} inflation \cite{chalo,chalo2,tsu}.}
\end{itemize}

As indicated above eq.(\ref{rangok}), the cosmologically relevant 
fluctuations have at the beginning of inflation physical
wavelengths in a range reaching the Planck scale 
$$ 
3.3  \; 10^{-32} \; e^{64-N_{tot}} \; \beta  \; {\rm cm} 
\lesssim \lambda^{init}
= 2 \, \pi/k^{init} \lesssim 3.3  \; 10^{-28} \; e^{64-N_{tot}} \; \beta 
\; {\rm cm} \; ,
$$
These fluctuations become macroscopic 
through the huge redshift during inflation and the subsequent expansion of
the universe with wavelengths today in the range 
$ 1 \, {\rm Mpc} \lesssim \lambda_{today} \lesssim 10^4 $ Mpc.
Namely, a total redshift of $ 10^{56} $. During this process these quantum
fluctuations classicalize just due to the huge stretching of the lengths. 
A field theoretical treatment shows that the quantum density matrix of the
inflaton becomes diagonal in the inflaton field representation as 
inflation ends \cite{chalo2}. 

\subsection{Slow-roll, the Universal Form of the Inflaton Potential 
and the Energy Scale of Inflation} \label{potuniv}

The inflaton potential $ V(\varphi) $ must be a slowly varying
function of $ \varphi $ in order to permit a slow-roll solution for
the inflaton field $ \varphi(t) $ which guarantees a total number of 
efolds $ \sim 64 $ as discussed in \cite{nos}.
Slow-roll inflation corresponds to a fairly flat potential and the
slow-roll approximation usually invokes a hierarchy of dimensionless
ratios in terms of the derivatives of the potential 
\cite{libros,hu}.
We recasted the slow-roll approximation as an expansion in $ 1/N $ 
where $ N \sim 60 $ is the number of efolds since the cosmologically 
relevant modes exit the horizon till the end of inflation \cite{1sN}.

In the slow-roll regime higher time derivatives can be neglected in the
evolution eqs.(\ref{eqno}) and (\ref{frinf}) with the result
\be\label{sr1}
3 \, H(t) \; {\dot \varphi} + V'(\varphi) = 0 \quad ,  \quad 
H^2(t) = \frac{V(\varphi)}{3 M^2_{Pl}} 
\ee
These first order equations can be solved in closed from as
\be \label{Nefo}
M^2_{Pl} \; N[\varphi] = -\int_{\varphi}^{\varphi_{end}}  \;
V(\varphi) \; \frac{d \psi}{dV} \; d\psi \;  \; .
\ee
where $ N[\varphi] $ is the number 
of e-folds since the field $ \varphi $ exits the 
horizon till the end of inflation (where $ \varphi $ takes the value 
$ \varphi_{end} $). This is in fact the slow roll solution 
of the evolution equations eqs.(\ref{eqno}) and (\ref{frinf}) 
in terms of quadratures.

Eq.(\ref{Nefo}) indicates that $ M^2_{Pl} \; N[\varphi] $ scales as $ \varphi^2 $ and therefore
the field $ \varphi $ is
of the order $ \sqrt{N} \; M_{Pl}\sim \sqrt{60}  \; M_{Pl} $ for the cosmologically relevant modes.
Therefore, we propose as universal form for the inflaton potential \cite{1sN}
\be \label{V} 
V(\varphi) = N \; M^4 \; w(\chi)  \; ,
\ee  
\noindent where $ \chi $ is a dimensionless, slowly varying field 
\be\label{chifla} 
\chi = \frac{\varphi}{\sqrt{N} \;  M_{Pl}}  \; ,
\ee 
More precisely, we choose $ N \equiv N[\varphi] $ as the number of e-folds since a 
pivot mode $ k_0 $ exits the horizon till the end of inflation.
Eq.(\ref{V}) includes all well known slow-roll families of inflation models such as new 
inflation \cite{infnue}, chaotic inflation \cite{infcao}, natural inflation \cite{natu}, etc.

\medskip

The dynamics of the rescaled field $ \chi $ exhibits the slow time 
evolution in terms of the \emph{stretched} dimensionless cosmic time 
variable, 
\be \label{tau} 
\tau =  \frac{t \; M^2}{M_{Pl} \; \sqrt{N}}  \quad , \quad  
{\cal H} \equiv \frac{H \; M_{Pl}}{\sqrt{N} \; M^2} = {\cal O}(1) \; .
\ee 
The rescaled variables $ \chi $ and $ \tau $ change slowly with time. 
Eq.(\ref{chifla}) shows that a
large change in the field amplitude $ \varphi $ results in a small 
change in the $ \chi $ amplitude; a change in $ \varphi \sim  M_{Pl} $ 
results in a $ \chi $ change $ \sim 1/\sqrt{N} $. The form of the 
potential, eq.(\ref{V}), the rescaled dimensionless inflaton field 
eq.(\ref{chifla}) and the time variable $ \tau $ make {\bf manifest} the 
slow-roll expansion as a consistent 
systematic expansion in powers of $ 1/N $ \cite{1sN}.  

We can choose $ |w''(0)| = 1 $ without loosing generality. Then, 
the inflaton mass scale around zero field is given by a see-saw formula
\be \label{m}
m^2  = \left| V''(\varphi=0) \right| = 
\frac{M^4}{M_{Pl}^2}  \quad ,  \quad m = \frac{M^2}{M_{Pl}} \; .
\ee
The Hubble parameter when the cosmologically relevant modes exit the 
horizon is given by
\be \label{hub}
H  = \sqrt{N} \; m \, {\cal H} \sim 7 \; m \; ,  
\ee
where we used that $  {\cal H} \sim 1 $. As a result, $ m\ll M $ and 
$ H \ll M_{Pl} $. The value of $ M $ is determined by the amplitude of the
CMB fluctuations within the effective theory of inflation. We obtain in 
sec. \ref{escene} [see eqs.(\ref{valorM}) and (\ref{myH})]: $ M \sim 0.70
\; 10^{16} $ GeV, $ m \sim 2.04 \; 10^{13} $ GeV and $ H \sim 10^{14} $ 
GeV for generic slow-roll potentials eq.(\ref{V}).

The energy density and the pressure [eq.(\ref{enerpres})] in terms of the 
dimensionless rescaled field $ \chi $ and the slow time variable $ \tau $ 
take the form,
\be\label{enepre2} 
\frac{\rho}{ N \; M^4} = \frac1{2\;N} \left(\frac{d\chi}{d \tau}\right)^2 
+ w(\chi) \quad ,\quad 
\frac{p}{ N \; M^4} = \frac1{2\;N} \left(\frac{d\chi}{d \tau}\right)^2 - 
w(\chi) \; . 
\ee
The equations of motion (\ref{eqno}) and (\ref{frinf}), in the same 
variables become
\bea \label{evol} 
&&  {\cal H}^2(\tau) = \frac{\rho}{ N \; M^4} =
\frac13\left[\frac1{2\;N} 
\left(\frac{d\chi}{d \tau}\right)^2 + w(\chi) \right] \quad , \cr \cr
&& \frac1{N} \;  \frac{d^2
\chi}{d \tau^2} + 3 \;  {\cal H} \; \frac{d\chi}{d \tau} + w'(\chi) = 0 
\quad .
\eea 
The slow-roll approximation follows by neglecting the
$ 1/N $ terms in eqs.(\ref{evol}). Both
$ w(\chi) $ and $ {\cal H}(\tau) $ are of order $ N^0 $ for large $ N $. 
Both equations make manifest the slow-roll expansion as an expansion in 
$ 1/N $.

Eq.(\ref{Nefo}) in terms of the field $ \chi $ takes the form
\be \label{Nchi}
-\int_{\chi_{exit}}^{\chi_{end}}  \; \frac{w(\chi)}{w'(\chi)} \; d\chi = 1 
\; .
\ee
This gives $ \chi= \chi_{exit} $ at horizon exit as a function of the couplings 
in the inflaton potential $ w(\chi) $.

Inflation ends after a finite number of efolds provided [see eq.(\ref{vmin})]
\be\label{condw}
w(\chi_{end}) = w'(\chi_{end}) = 0 \; .
\ee
So, this condition is enforced in the inflationary potentials.

\medskip

\noindent
For the quartic degree potentials $ V(\varphi) $ 
eqs.(\ref{nueva})-(\ref{caotica}), the corresponding dimensionless 
potentials $ w(\chi) $ take the form
\bea\label{wnue}
&& w(\chi) = \frac{y}{32} \left(\chi^2 - \frac8{y}\right)^{\! 2} 
= -\frac12 \; \chi^2 
+ \frac{y}{32} \; \chi^4 + \frac2{y}\quad , \quad {\rm new \; inflation} \; , \\ \cr 
&& w(\chi) = \frac12 \; \chi^2 + \frac{y}{32} \; \chi^4 \quad , 
\quad {\rm chaotic \; inflation} \; , \label{wchao}
\eea
where the coupling $ y $ is of {\bf order one} and 
$$
\lambda = \frac{y}{8 \; N} \left( \frac{M}{M_{Pl}}\right)^4 \ll 1 \quad 
{\rm since} \quad  M \ll  M_{Pl} \; .
$$
For a general potential $ V(\varphi) $ we can always eliminate the linear 
term by a shift in the field $ \varphi $ without loosing generality,
\be \label{serie}
V(\varphi) = V_0 \pm \frac12 \; m^2 \; \varphi^2 + 
\sum_{n=3}^{\infty} \frac{\lambda_n}{n} \; \varphi^n \; ,
\ee
and 
\be\label{seriew}
w(\chi) = w_0 \pm \frac12 \; \chi^2 + \sum_{n=3}^{\infty} 
\frac{G_n}{n} \; \chi^n \; ,
\ee
where the dimensionless coefficients $ G_n $ are of order one.
We find from eqs. (\ref{V}) and (\ref{chifla}),
\be\label{Gn}
 V_0 = N \; M^4 \;  w_0 \quad , \quad 
\lambda_n =  \frac{G_n \; M^4}{N^{\frac{n}2-1} \; M_{Pl}^n} \; ,
\ee
In particular, we get comparing with eqs.(\ref{nueva}), (\ref{caotica}),
(\ref{wnue}) and (\ref{wchao}),
\be\label{lambdaG4}
\lambda_3 = \frac{G_3}{\sqrt{N}}
\; \frac{M^4}{M_{Pl}^3}  \quad ,  \quad \lambda = \lambda_4  = 
\frac{G_4}{N} \left(\frac{M}{M_{Pl}}\right)^4 \quad , \quad G_4 = 
\frac{y}8 \; .
\ee
We find the dimensionful couplings $ \lambda_n $ suppressed by the nth 
power of $ M_{Pl} $
as well as by the factor $ N^{\frac{n}2-1} $. Notice that this suppression 
factors are natural and come from the ratio of the two relevant energy 
scales here: the Planck mass and the inflation scale $ M $.

\medskip

In new inflation with the potential of eq.(\ref{wnue}), the inflaton 
starts near the local maximum $ \chi = 0 $ and keeps rolling down the 
potential hill till it reaches the absolute minimum $ \chi = \sqrt{8/y} $. 
The initial values of $ \chi $
and $ \dot\chi $ must be chosen to have a total of $ \gtrsim 64 $ efolds of
inflation. In all cases $ \chi(0) $ and $ \dot\chi(0) $ turn to be of order
one.

\medskip

There are two {\it generic} inflationary regimes: slow-roll and fast-roll 
depending on whether
\bea
&& \frac1{2\;N} \left(\frac{d\chi}{d \tau}\right)^2 \ll w(\chi) \quad  : 
\quad {\rm  slow-roll \; regime} \cr \cr
&& \frac1{2\;N} \left(\frac{d\chi}{d \tau}\right)^2 \sim w(\chi) \quad  : 
\quad {\rm  fast-roll \; regime}
\eea
Both regimes show up in {\bf all} inflationary models in the class 
eq.(\ref{V}). A fast-roll stage emerges from generic initial 
conditions for the inflaton field. This fast-roll stage is generally  very
short and is followed by the slow-roll stage (see sec. \cite{nos,singu}).
The slow-roll regime is an attractor in this dynamical system \cite{bgzk}.

\medskip

Eq.(\ref{V}) for the inflaton potential resembles the
moduli potential coming from supersymmetry breaking,
\be\label{susy}
V_{susy}(\varphi) =  
m_{susy}^4 \; v\!\left(\frac{\varphi}{M_{Pl}}\right) \; ,
\ee
where $ m_{susy} $ stands for the supersymmetry breaking scale. 
In our context, eq.(\ref{susy}) implies that 
$  m_{susy} \sim 10^{16}$ Gev. That is, the supersymmetry breaking scale
$ m_{susy} $ turns out to be at the GUT scale $ m_{susy} \sim M_{GUT} $.

It must be stressed that the validity of the inflaton potential 
eq.(\ref{V}) is independent of whether or not there is an underlying 
supersymmetry. In addition, the observational support on inflaton 
potentials like eq.(\ref{V}) can be taken as a first signal 
of the presence of supersymmetry in a cosmological context.
No experimental signals of supersymmetry are known so far despite
the enormous theoretical work done on supersymmetry since 1971.

\begin{figure}[h]
\begin{center}
\begin{turn}{-90}
\psfrag{"lna.dat"}{$ \ln a(\tau) $ vs. $ \tau $}
\includegraphics[height=10cm,width=7cm]{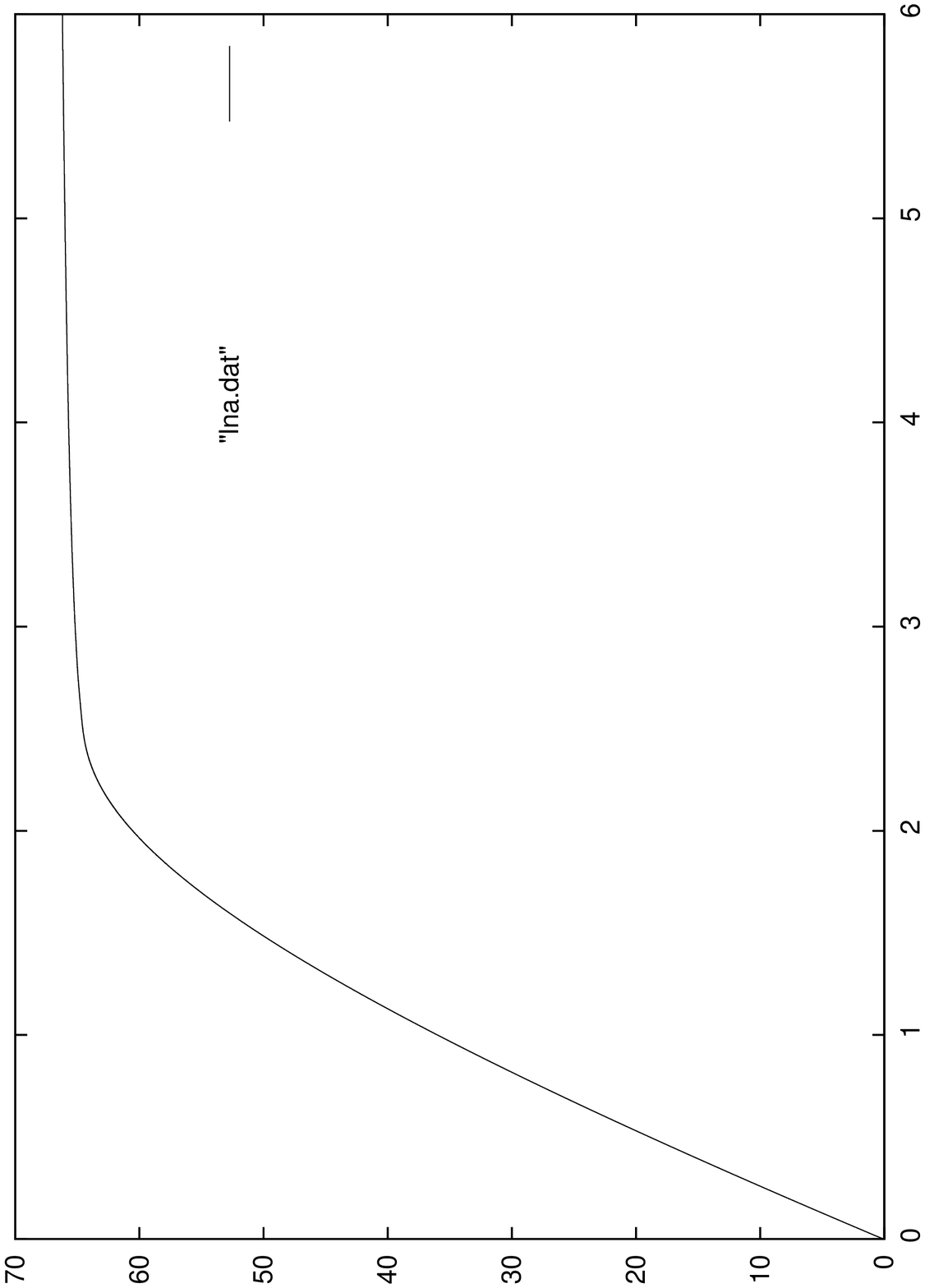}
\psfrag{"pe.dat"}{$ p/\rho $ vs. $ \tau $}
\includegraphics[height=10cm,width=7cm]{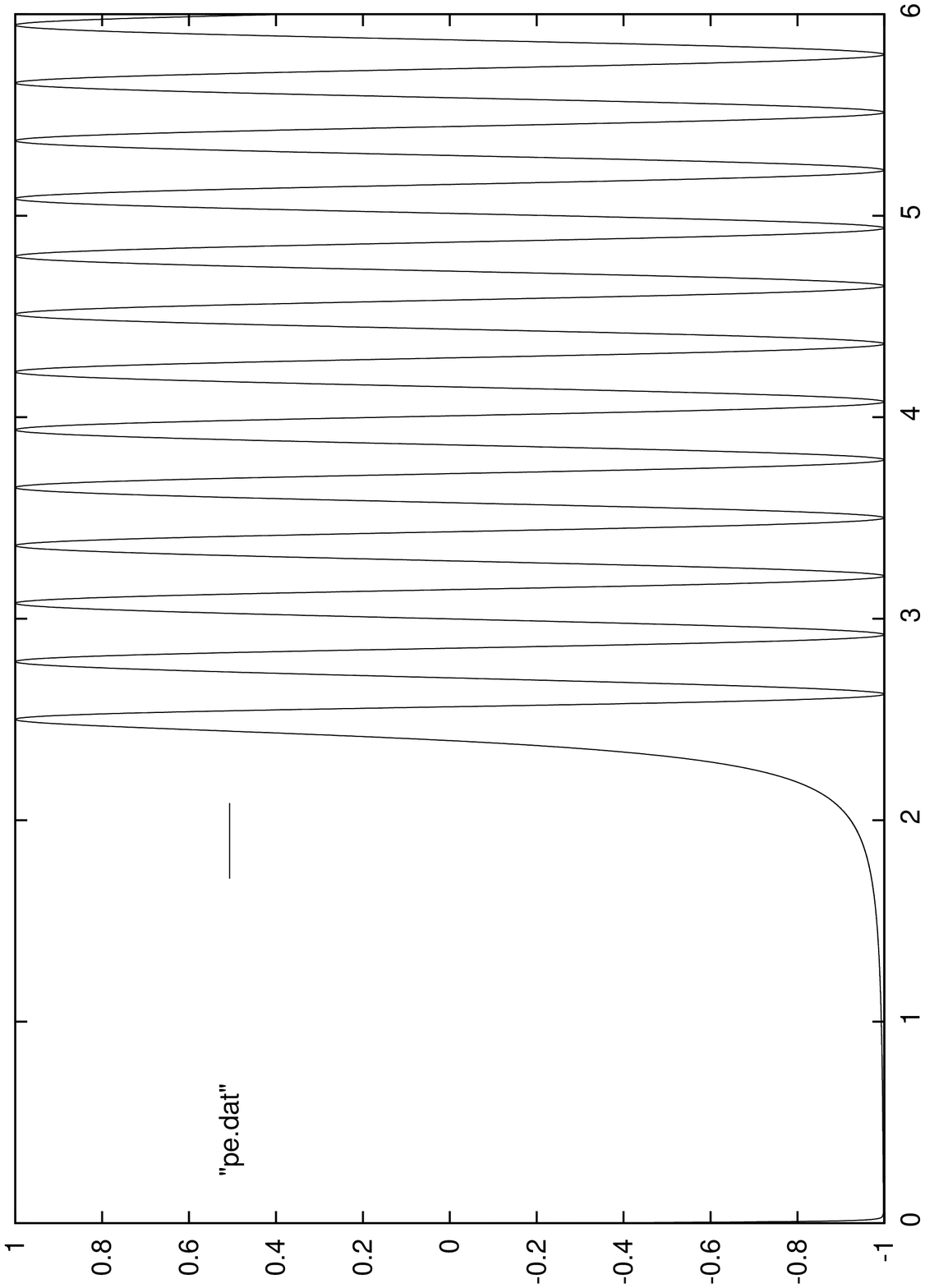}
\end{turn}
\caption{Upper panel: $ \ln a(\tau) $ which is the number of e-folds
as a function of the stretched cosmic time eq.(\ref{tau}). 
We see that $ a(\tau) $ grows exponentially with time
(quasi-de Sitter inflation) for $ \tau < \tau_{end} \simeq 2.39 $.
Lower panel: the equation of state  $  p / \rho $ vs. $ \tau $ 
[eq.(\ref{eqnofstate})]. We have $ p / \rho \simeq - 1 $ after the 
fast-roll stage and before the end of inflation. That is, for
$ 0.0247 \lesssim\tau <   \tau_{end} \simeq 2.39 . \;  p / \rho $ 
oscillates with zero average: $ <p / \rho> = 0 $ during the subsequent 
matter dominated era. Both figures are for the inflaton potential
eq.(\ref{wnue}) with $ y = 1.26 $ and $ N_{tot} = 64 $ efolds of 
inflation.}
\label{infla1}
\end{center}
\end{figure}

\begin{figure}[h]
\begin{center}
\begin{turn}{-90}
\psfrag{"fi.dat"}{$ \chi(\tau) $ vs. $ \tau $}
\psfrag{"fip.dat"}{$ {\dot \chi}(\tau) $ vs. $ \tau $}
\includegraphics[height=10cm,width=7cm,keepaspectratio=true]{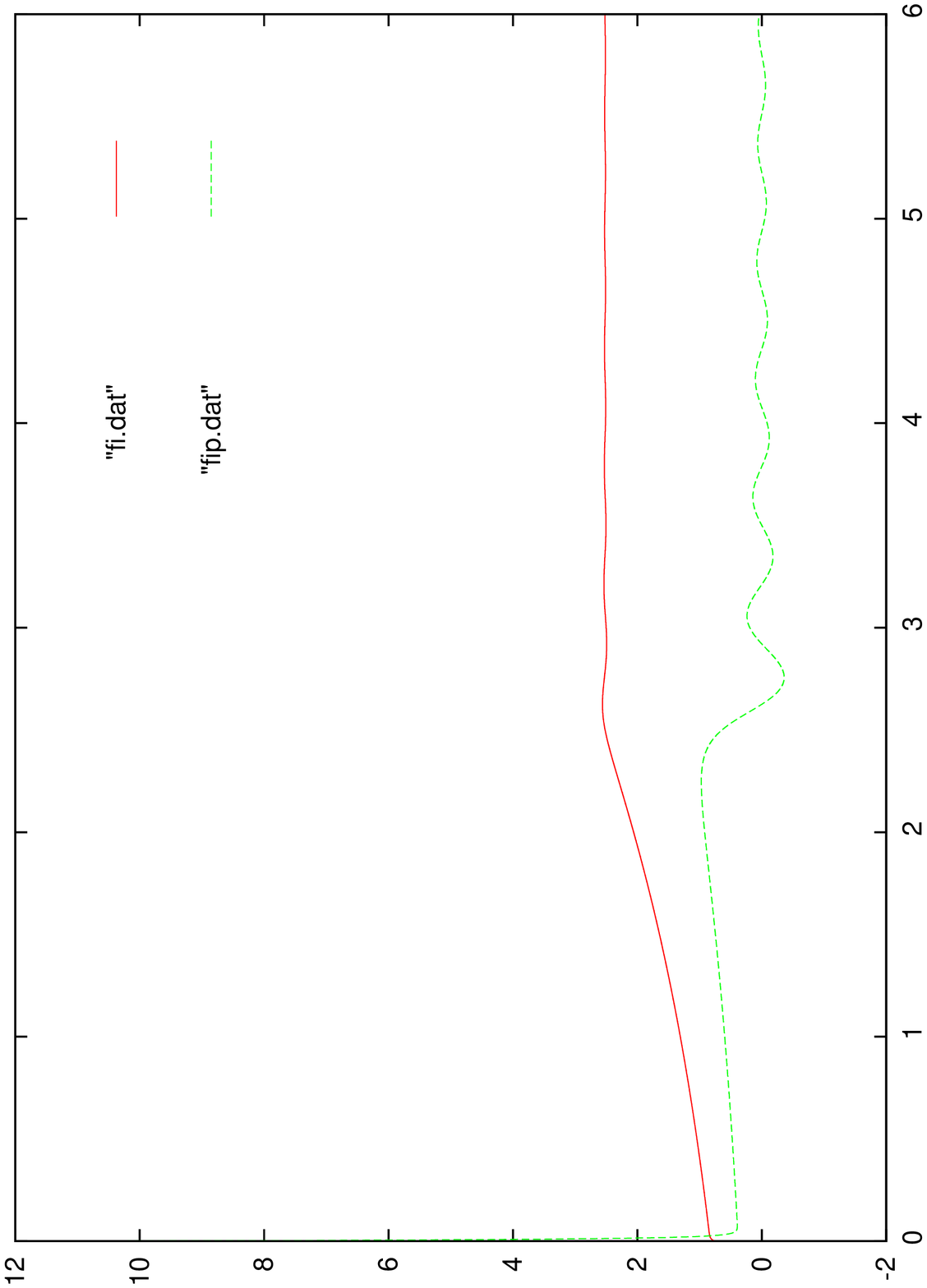}
\psfrag{"ush.dat"}{ $ 1/{\cal H}(\tau) $ vs. $ \tau $}
\includegraphics[height=10cm,width=7cm,keepaspectratio=true]{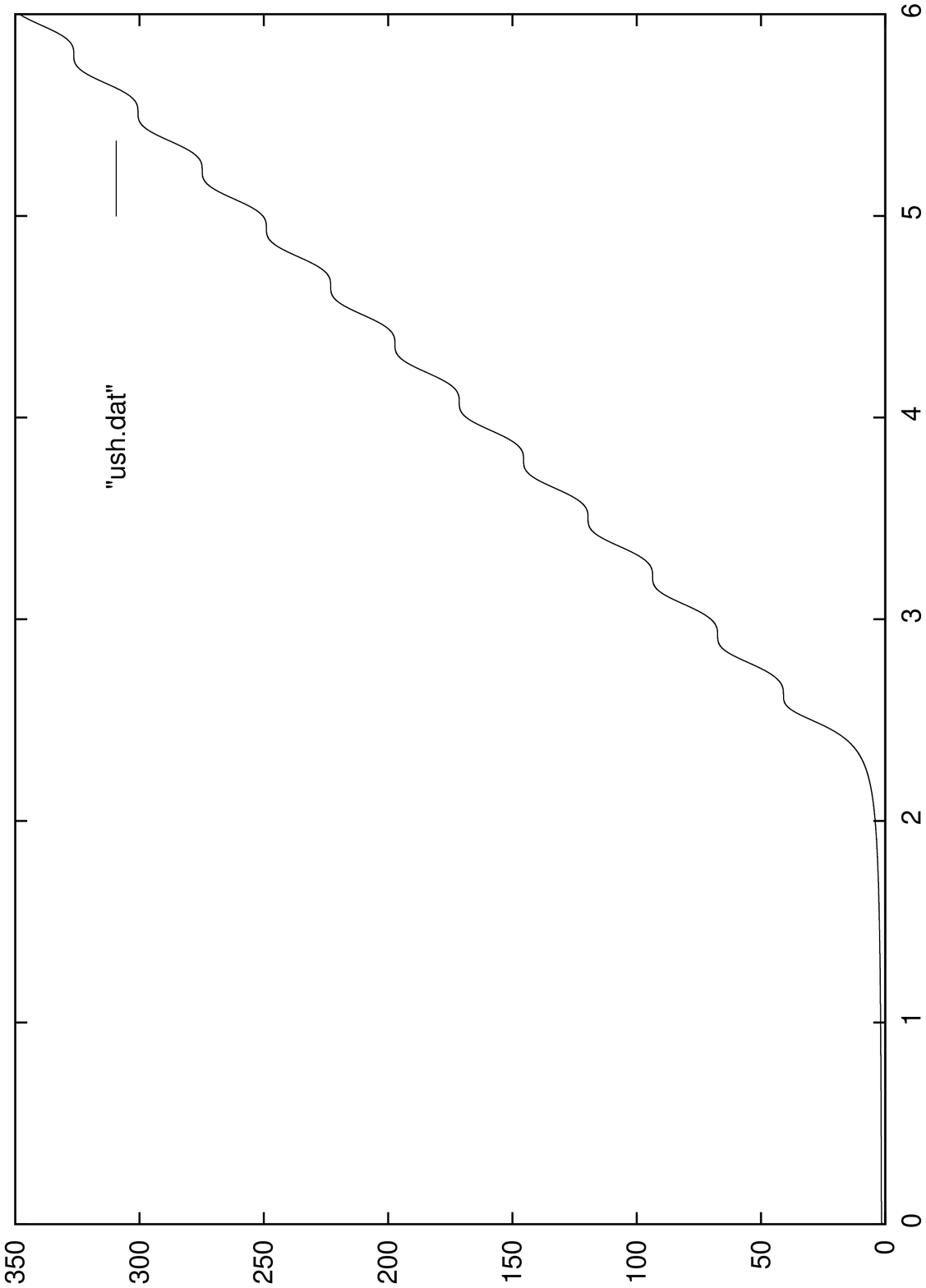}
\end{turn}
\caption{Upper panel: $ \chi(\tau) $ and  $ \dot\chi(\tau)$
as a function of the stretched  cosmic time $ \tau $ for $ \chi(0) = 
0.740 $ and initial kinetic energy equal to the initial potential energy 
which implies $ {\dot \chi}(0) = 12.6 $. After a short 
fast-roll stage for $ \tau\lesssim 0.0247 $ 
the inflaton field slowly rolls toward its absolute minimum at 
$ \chi = \sqrt{8/y} \simeq 2.52\ldots, \; { \dot\chi } = 0 $. 
Lower panel: $ 1/{ \cal H} $ vs. $ \tau $.
$ 1/{ \cal H} $ grows slowly during inflation $ \tau < \tau_{end} \simeq 
2.39 $ and grows as $ 1/{\cal H} \simeq \frac32 \; N \; (\tau-\tau_{end}) $
in the subsequent matter dominated era.}
\label{infla2}
\end{center}
\end{figure}

\subsection{The energy scale of inflation and the quasi-scale
invariance during inflation.}\label{escene}

The inflationary scalar power spectrum can be written as \cite{hu,nos}
\be  \label{potBD}
P^{BD}_\mathcal{R}(k) = |{\Delta}_{k\;ad}^{\mathcal{R}}|^2 \; 
\left(\frac{k}{k_0}\right)^{n_s - 1} \; , 
\ee 
where in the slow-roll approximation (leading order in $ 1/N $)
the amplitude 
$ |{\Delta}_{k\;ad}^{\mathcal{R}}|^2 $  takes the form \cite{nos}
\be \label{ampliI}
|{\Delta}_{k\;ad}^{\mathcal{R}}|^2  = \frac{N^2}{12 \, \pi^2} \;
\left(\frac{M}{M_{Pl}}\right)^4 \; \frac{w^3(\chi)}{{w'}^2(\chi)} \; .
\ee
where $ \chi \equiv \chi_{exit} $ stands for the inflaton field at 
horizon exit and $ n_s $ stands for the spectral index
\be\label{ns}
 n_s - 1 = -\frac3{N} \; \left[\frac{w'(\chi)}{w(\chi)} \right]^2
+  \frac2{N}  \; \frac{w''(\chi)}{w(\chi)} +
{\cal O}\left(\frac1{N^2}\right)\quad .
\ee
Since, $ w(\chi) $ and  $ w'(\chi) $ are of order one, 
we find from eq.(\ref{ampliI})
\be\label{Mwmap}
\left(\frac{M}{M_{Pl}}\right)^2 \sim \frac{2
\, \sqrt3 \, \pi}{N} \; |{\Delta}_{k\;ad}^{\mathcal{R}}| \simeq  0.897
\times 10^{-5} \; .
\ee
where we used $ N \simeq 60 $, set $ k = k_0 $ with  
$ k_0 = 0.002 \; ({\rm Mpc})^{-1} $ the WMAP pivot scale and ref.\cite{WMAP5}
\be\label{aniso}
|{\Delta}_{k\;ad}^{\mathcal{R}}| = (4.94 \pm 0.1)\times 10^{-5} \; .
\ee
This fixes the scale of inflation to be
\be\label{valorM}
M \sim 2.99 \times 10^{-3} \; M_{Pl} \sim 0.73
\times 10^{16}\,\textrm{GeV} \; .
\ee
This value {\em pinpoints the scale of the potential} during inflation 
to be at the GUT scale suggesting a deep connection between inflation 
and the physics at the GUT scale in cosmological space-time.

As a consequence we get for the inflaton mass and
the Hubble parameter during inflation from eq.(\ref{m})-(\ref{hub}),
\be\label{myH}
m = \frac{M^2}{M_{Pl}} \sim 2.18 \times 10^{13} \, \textrm{GeV} 
\quad , \quad H \sim 10^{14} \, \textrm{GeV} 
\ee
Notice that these values for the inflation scale $ M $ and the inflaton 
mass are {\bf model independent} within the slow-roll class of models 
eq.(\ref{V}). In addition, we see that $ m \simeq 0.003 \; M $. 
Namely, the inflaton is a {\bf very light} field in this context.
We can therefore expect infrared and scale invariant phenomena here.

Since $ M/M_{Pl} \sim 3 \times 10^{-3} $ [eq.(\ref{valorM})],
we {\bf naturally} find from eq.(\ref{lambdaG4}) the order of
magnitude of the cubic and quartic couplings, 
\be  \label{natu} 
\lambda_3 \sim 10^{-6} \; m \quad ,  \quad \lambda=\lambda_4 
\sim 10^{-12} \quad . 
\ee 
These relations are a {\bf natural} consequence
of the validity of the effective field theory and of slow-roll and
solve the {\bf fine tuning problem}.  We emphasize that the
`see-saw-like' form of the couplings is a consequence of the form of the 
potential eq.(\ref{V}) and is valid for all inflationary models within
the class defined by eq.(\ref{V}) \cite{1sN}.

We obtain from at the coupling value $ y = 1.26 $ that 
best fit the WMAP5+SDSS+SN data \cite{mcmc,nos}
\be \label{myM}
M = 0.543 \times 10^{16} \;  {\rm GeV} \quad ,  \quad
m = 1.21 \times 10^{13} \; {\rm GeV ~~~ and} \quad H \simeq 6 \times  
10^{13} {\rm ~GeV ~~~ for}  \quad   y = 1.26  \; . 
\ee
Notice that these values {\bf agree} with the generic estimates 
eq.(\ref{valorM})-(\ref{myH}) which apply in order of magnitude
to all inflationary potentials within the universal slow-roll class 
eq.(\ref{V}). 

\medskip

The ratio $ r $ to leading order in $ 1/N $ is given by \cite{nos}
\be \label{indi}
r = \frac8{N} \; \left[\frac{w'(\chi)}{w(\chi)} \right]^2 +
{\cal O}\left(\frac1{N^2}\right)\quad .
\ee
The fact that $ r \sim 1/N $ [eq.(\ref{indi})] shows that
the tensor fluctuations are suppressed by a factor $ N \sim 60 $ 
compared with the curvature scalar fluctuations.
This suppression can be explained as follows: 
the scalar curvature fluctuations are quantum 
fluctuations around the classical inflaton while the tensor 
fluctuations are just quantum zero-point fluctuations.

\medskip

The observation of a nonzero $ r $ will have {\bf twofold} relevance.
First, it would be the {\bf first} detection of (linearized)
gravitational waves as predicted by Einstein's General Relativity.
Second, $ r > 0 $ indicates the presence of gravitons,
namely, {\bf quantized} gravitational waves at tree level.

\medskip

Neutrino oscillations and neutrino masses $ m_{\nu} $ 
are currently explained in the see-saw mechanism as follows \cite{ita},
\be\label{neum}
\Delta m_{\nu} \sim \frac{M^2_{Fermi}}{M_R} 
\ee
where $ M_{Fermi} \sim 250$ GeV is the Fermi mass scale, 
$ M_R \gg  M_{Fermi} $ is a large energy scale and $ \Delta m_{\nu} $ 
is the difference between the neutrino masses for different flavors. 
The observed values for $ \Delta m_{\nu} \sim 0.009 - 0.05 $ eV 
naturally call for a mass scale $ M \sim 10^{15-16}$ GeV close to 
the GUT scale \cite{ita}. 

We see thus, that the energy scale $ \sim 10^{16}$ GeV
appears in fundamental physics in at least three independent ways:
grand unification scale, inflation scale and the scale $ M_R $
in the see-saw neutrino mass formula.

\subsection{Ginsburg-Landau polynomial realizations of the Inflaton 
Potential}

In the Ginsburg-Landau spirit the potential is a polynomial in the field
starting by a constant term \cite{gl}. Linear terms can always be 
eliminated by a constant shift of the inflaton field. The quadratic term 
can have a positive or a negative sign. In the first case 
the symmetry $ \varphi \to - \varphi $ is unbroken (unless the potential 
contains terms odd in $ \varphi $), in the latter case  the symmetry 
$ \varphi \to - \varphi $ is spontaneously broken since 
the minimum of the potential is at $ \varphi \neq 0 $. 

Inflaton potentials with $ w''(0) > 0 $ lead to chaotic (large field) 
inflation while inflaton potentials with $ w''(0) < 0 $ lead to new (small
field) inflation.

The inflaton potential must be bounded from below, therefore the next potential beyond
the quadratic potential is the quartic one with a positive quartic coefficient.

The request of renormalizability restricts the degree of the inflaton potential to four.
However, since the theory of inflation is an effective theory,  potentials of degrees
higher than four are acceptable.

In the context of the Ginsburg-Landau effective theory it is highly 
unnatural to set $ m=0 $ \cite{gl}. This corresponds to be exactly at the critical point 
of the model where the mass vanishes, that is,  in the statistical mechanical context
the correlation length is infinite. In fact, the WMAP result convincingly excluding 
the $ m = 0 $ choice (purely $ \varphi^4 $ potential, see \cite{nos,WMAP1,WMAP5})
supports this purely theoretical argument against the $ \varphi^4 $ monomial potential.
Therefore, from a physical  point of view, the pure quartic potential
is a weird choice implying to fine tune to zero the coefficient of the mass term. 

\medskip 

Dropping the cubic term implies that $ \varphi \to -\varphi $ is a symmetry of the 
inflaton potential. As stated in \cite{ciri},
we do not see reasons based on fundamental physics to choose a zero or a 
nonzero cubic term. However, the MCMC analysis of the WMAP plus LSS data 
shows that the cubic term can be ignored for new inflation (see \cite{mcmc}). 

\medskip 

A model with only one field is clearly unrealistic since the inflaton would 
then describe a stable and ultra-heavy ($ \sim 10^{13}$GeV) particle. It is
necessary to couple the inflaton with lighter particles, in which case the inflaton
can decay into them. 
There are many available scenarios for inflation. Most of them add
other fields coupled to the inflaton. This variety of inflationary
scenarios may seem confusing since several of them are compatible with the
observational data \cite{WMAP1,WMAP5}. Indeed, future observations should
constraint the models more tightly excluding some families of them. 
The hybrid inflationary \cite{hibri} models are amongst those 
strongly disfavoured by
the WMAP data since they give $ n_s > 1 $ in most of their parameter space 
contrary to the WMAP results \cite{WMAP5}.
The regions of parameter space where hybrid inflation yields $ n_s < 1 $
are equivalently covered by one-field chaotic inflation \cite{infwmap}.

The variety of inflationary models shows the {\bf power} of the
inflationary paradigm. Whatever the correct microscopic model for
the early universe would be, it should include inflation with the generic 
features we know today. Many inflatons can be considered 
(multi-field inflation), but such family of models introduce extra 
features as non-adiabatic (isocurvature) density fluctuations, 
which in turn become strongly constrained by the WMAP 
data \cite{WMAP1,WMAP5}.

\section{Overview of Dark Matter}\label{secdm}

A {\bf model independent} analysis of dark matter (DM) both 
decoupling ultrarelativistic (UR) and non-relativistic (NR) based on the
DM phase-space density $ \mathcal{D} = \rho_{DM}/\sigma^3_{DM} $
is presented in refs. \cite{oscnos,mnras,dsg}. We derived in \cite{mnras} explicit 
formulas for the DM particle mass $ m $ and for the number of 
ultrarelativistic degrees of freedom $ g_d $ at decoupling
(and equivalently for the decoupling temperature $ T_d $). 
We find that for DM particles decoupling UR both at local 
thermal equilibrium (LTE) and out of LTE, $ m $ turns to be at the 
{\bf keV scale}. For example, for DM Majorana fermions
decoupling at LTE the mass results $ m \simeq 0.85 $ keV.
For DM particles decoupling NR, $ \sqrt{m \; T_d} $ results in the 
{\bf keV scale} ($ T_d $ is the decoupling temperature) and the 
$ m $ value is consistent with the keV scale. 
Also, lower and upper bounds on the DM annihilation 
cross-section for NR decoupling are derived.
We evaluate the free-streaming (Jeans') wavelength and Jeans' mass: they 
result independent of the type of DM except for the DM self-gravity 
dynamics. The free-streaming wavelength today results in the  
{\bf kpc range}. These results are based on our 
theoretical analysis, astronomical observations of dwarf spheroidal 
satellite galaxies in the Milky Way and $N$-body numerical simulations.
We analyze and discuss the results on $ \mathcal{D} $ from analytic 
approximate formulas both for linear fluctuations and the (non-linear) 
spherical model and from $N$-body simulations results. We obtain in this 
way upper bounds for the DM particle mass which all result below the 100 
keV range \cite{mnras}.

\subsection{The dark matter particle mass}

Dark matter constitutes 83 \% of the matter in the Universe.
Its nature is still unknown. The external part of galaxies (halo) is formed
by dark matter. Baryonic matter dominates the internal parts of luminous
galaxies.

Dark matter (DM) must be non-relativistic by the time of 
structure formation ($ z < 30 $) in order to reproduce the observed small 
structure at $ \sim 2-3 $ kpc. 

DM particles can decouple being 
ultrarelativistic (UR) at $ \; T_d \gg m $ or being non-relativistic (NR) 
at $ \; T_d \ll m $ where $ m $ is the mass of the DM particles 
and $ T_d $ the decoupling temperature. 
We consider DM particles that decouple {\bf at} or {\bf out} of local thermal 
equilibrium (LTE) \cite{oscnos,mnras}.

The DM distribution function $ F_d $ freezes out at decoupling. Therefore,
for all times after decoupling $ F_d $ coincides 
with its expression at decoupling. $ F_d $ is a function of $ T_d, \; m $ 
and the comoving momentum of the DM particles $ p_c $.

Knowing the distribution function $ F_d(p_c) $, we can compute physical 
magnitudes as the DM velocity fluctuations and the DM energy density. 
For the relevant times $ t $ during structure formation, when the DM 
particles are non-relativistic, we have
\be\label{flucvel}
\langle \vec{V}^2\rangle(t) = \langle \frac{\displaystyle 
\vec{p}^{\,2}_{ph}}{m^2} 
\rangle(t) = \frac{\displaystyle \int \frac{d^3 p_{ph}}{(2\pi)^3}  \;  
\frac{\vec{p}^{\,2}_{ph}}{m^2} \; 
F_d[a(t) \, p_{ph}]}{\displaystyle \int \frac{d^3p_{ph}}{(2\pi)^3} \; 
F_d[a(t) \; p_{ph}]}
\ee
where we use the physical momentum of the DM particles 
$ p_{ph}(t) \equiv p_c/a(t) $ as integration variable. 
The physical momentum $ p_{ph}(t) $ coincides today with the comoving 
momentum $ p_c $.

We can relate the covariant decoupling temperature $ T_d $, the effective
number of UR degrees of freedom at decoupling $ g_d $ and the photon 
temperature today $ T_{\gamma} $ by using entropy conservation 
\cite{kt,gb,pdg}: 
\be\label{temp}
T_d = \left(\frac2{g_d}\right)^\frac13 \; T_{\gamma} \; 
, \quad {\rm where} \quad T_{\gamma} = 0.2348 \; {\rm meV} \quad {\rm and} \quad
1 \; {\rm meV} =  10^{-3} \; {\rm eV} \; .
\ee
The DM energy density can be written as
\be \label{roDM}
\rho_{DM}(t) = g\int \frac{d^3p_{ph}}{(2\pi)^3} \; \sqrt{m^2+p^2_{ph}} ~ 
F_d[a(t) \, p_{ph}] \; ,
\ee
where $ g $ is the number of internal degrees of freedom of the DM 
particle, typically $ 1 \leq g \leq 4 $.

By the time when the DM particles are non-relativistic, the
energy density eq.(\ref{roDM}) becomes
\be \label{roDM2}
\rho_{DM}(t) = \frac{m~g}{2 \; \pi^2} \; \frac{T^3_d}{a^3(t)} \; I_2 
\equiv m \; n(t) \; , 
\ee
where
$$
I_2 \equiv \int_0^{\infty} y^2 \; F_d(y) \; dy \; ,
$$
$ n(t) $ is the number of DM particles per unit volume and 
we used as integration variable 
\be\label{varin}
y \equiv \frac{p_{ph}(t)}{T_d(t)} = \frac{p_c}{T_d} \; . 
\ee
From eq.(\ref{roDM2}) at $ t = 0 $ and from the value observed today for 
$ \rho_{DM} $ [table I and eq.(\ref{rhocrit})], 
we find the value of the DM mass:
\be\label{mdm}
m = \pi^2 \; \Omega_{DM} \; \frac{\rho_c}{T_{\gamma}^3} \;
\frac{g_d}{g \; I_2}=
6.986 \; \mathrm{eV} \; \frac{g_d}{g \; I_2} \; , 
\ee
where $ \rho_c $ is the critical density.

Using as integration variable $ y $ [eq.(\ref{varin})], 
eq.(\ref{flucvel}) for the velocity fluctuations, yields
\be\label{v2}
\langle \vec{V}^2\rangle(t) = \left[\frac{T_d}{m \; a(t)}\right]^2  \; 
\frac{I_4}{I_2} \; , 
\ee
where
$$
I_4 \equiv \int_0^\infty y^4  \; F_d(y)  \; dy \; .
$$
Expressing $ T_d $ in terms of the CMB temperature today according to
eq.(\ref{temp}) gives for the one-dimensional velocity dispersion,
\be\label{sigdm}
\sigma_{DM}(z) = \sqrt{\frac13 \; \langle \vec{V}^2 \rangle(z)} = 
\displaystyle \frac{2^{\frac13}}{\sqrt3} \; \frac{1+z}{g^{\frac13}_d} \; 
\frac{T_{\gamma}}{m} \; \sqrt{\frac{I_4}{I_2}}=0.05124 \; \displaystyle 
\frac{1+z}{g^{\frac13}_d} \; \frac{\mathrm{keV}}{m}  \;
\left[\frac{I_4}{I_2}\right]^{\frac12} \; 
\frac{\mathrm{km}}{\mathrm{s}} \; \; .
\ee
It is very useful to consider the phase-space density invariant under the 
universe expansion \cite{oscnos,hogan}
\be \label{defD}
\mathcal{D}(t)  \equiv \frac{n(t)}{\langle \vec{P}^2_{ph}(t) 
\rangle^\frac32} \buildrel{\rm non-rel}\over= 
\frac1{3 \; \sqrt3  \; m^4} \; \frac{\rho_{DM}(t)}{\sigma^3_{DM}(t)}
\quad , 
\ee
where we consider the relevant times $ t $ during structure formation when
the DM particles are non-relativistic.
$ \mathcal{D}(t) $ is a \emph{constant} in absence of self-gravity. In the
non-relativistic regime $ \mathcal D(t) $ can only {\bf decrease} by
collisionless phase mixing or self-gravity dynamics \cite{theo}.

We derive a useful expression for the phase-space density $ {\mathcal D} $
from eqs.(\ref{roDM2}), (\ref{sigdm}) and (\ref{defD}) with the result 
\be \label{Dex} 
\displaystyle \mathcal{D} = \frac{g}{2 \; \pi^2}
\frac{I_2^{\frac52}}{I_4^{\frac32}}\; , 
\ee

\noindent
Observing dwarf spheroidal satellite galaxies in the Milky Way (dSphs) 
yields for the phase-space density today \cite{gilmore}:
\be\label{gil}
\frac{\rho_s}{\sigma^3_s} \sim 5\times 10^3 ~
\frac{\mathrm{keV}/\mathrm{cm}^3}{\left( \mathrm{km}/\mathrm{s}
\right)^3} = (0.18 \;  \mathrm{keV})^4 \; .
\ee
The precision of these results is about a factor $10$.

After the radiation dominated era the phase-space density reduces 
by a factor that we call $ Z $ 
\be\label{F1}
\mathcal{D}(0) = \frac1{Z} \; \mathcal{D}(z \sim 3200)
\ee
Recall that $ \mathcal{D}(z) $  [eq.(\ref{defD})] is independent of $ z $ for $ z \gtrsim 3200 $
since density fluctuations were $ \lesssim 10 ^{-3} $ before the matter dominated 
era \cite{libros}.

The range of values of $ Z $ (which is necessarily  $ Z > 1 $) 
is analyzed in detail in ref. \cite{mnras}.

We can express the phase-space density today from eqs.(\ref{defD}) and (\ref{gil}) as
\be\label{D0}
\mathcal{D}(0) = \frac1{3 \; \sqrt3  \; m^4} \; \frac{\rho_s}{\sigma^3_s} \; .
\ee
Therefore, eqs.(\ref{defD}), (\ref{F1}) and (\ref{D0}) yield,
\be\label{F}
\frac{\rho_s}{\sigma^3_s} = \frac1{Z} \; \frac{\rho_{DM}}{\sigma^3_{DM}}(z \sim 3200) \; ,
\ee
where $ \rho_{DM}/\sigma^3_{DM}(z \sim 3200) $ follows from eqs.(\ref{defD}) and (\ref{Dex}),
\be\label{denhoy}
\frac{\rho_{DM}}{\sigma^3_{DM}}(z \sim 3200) = \frac{3 \; \sqrt3  \; m^4}{2 \; \pi^2} \;
g \;\frac{I_2^{\frac52}}{I_4^{\frac32}} \; . 
\ee
We can express $ m $ from eqs.(\ref{gil})-(\ref{denhoy}) in terms of 
$ \mathcal{D} $ and observable quantities as
\bea\label{mD}
&&m^4 = \frac{Z}{3 \; \sqrt3} \; \frac{\rho_s}{\mathcal{D} \; \sigma^3_s}
= \frac{2 \; \pi^2}{3 \; \sqrt3} \; \frac{Z}{g} \; 
\frac{\rho_s}{\sigma^3_s} \; 
\frac{I_4^{\frac32}}{I_2^{\frac52}} \; , \\ \cr
&& m = 0.2504 \; \left(\frac{Z}{g}\right)^\frac14 \; \; 
\frac{I_4^{\frac38}}{I_2^{\frac58}} \; 
\mathrm{keV} \;  . \label{mDn}
\eea
Combining this with eq.(\ref{mdm}) for $ m $ we obtain the number of 
ultrarelativistic degrees of freedom at decoupling as
\be\label{gdD}
g_d = \frac{2^\frac14}{3^\frac38 \; \pi^\frac32} \; 
\frac{g^\frac34}{\Omega_{DM}} \; \frac{T_{\gamma}^3}{\rho_c} \; 
\left(\frac{Z \; \rho_s}{\sigma^3_s}\right)^\frac14 \;
\left[I_2 \; I_4\right]^{\frac38} = 35.96 \; Z^\frac14 \; g^\frac34 \; 
\left[I_2 \; I_4\right]^{\frac38} \; .
\ee
A succession of several violent phases happens during the structure 
formation 
stage ($ z \lesssim 30 $). Their cumulated effect together with the
evolution of $ \mathcal{D} $ for $ 3200 \gtrsim z \gtrsim 30 $ 
produces a range of values of the 
$ Z $ factor which we can conservatively estimate
on the basis of the $N$-body simulations results
\cite{numQ} and the approximation results from the linear
approximation and the spherical model \cite{mnras}.
This gives a range of values 
$ 1 < Z < 10000 $ for dSphs \cite{mnras}.

\subsection{Jeans'  (free-streaming)  
wavelength and Jeans' mass}\label{jeans}

It is very important to evaluate the Jeans' length and Jeans' mass in the 
present context \cite{gb,gilbert,bs}. The Jeans' length is analogous to 
the free-streaming wavelength. The free-streaming wavevector 
is the largest 
wavevector exhibiting gravitational instability and characterizes the 
scale of suppression of the DM transfer function during matter domination \cite{egil}.

The physical free-streaming wavelength can be expressed as \cite{gb,egil}
\be\label{fs}
\lambda_{fs}(t) = \lambda_J(t) = \frac{2 \; \pi}{k_{fs}(t)}
\ee
where $ k_{fs}(t) = k_J(t) $ is the  physical free-streaming wavenumber given by
\be\label{kfs}
k_{fs}^2(t) = \frac{4 \; \pi \; G \; \rho_{DM}(t)}{\langle 
\vec{V}^2\rangle(t)}= \frac32 \; [1+z(t)] \; \frac{H_0^2 \; 
\Omega_{DM}}{\langle \vec{V}^2\rangle(0)} \;  \; .
\ee
where we used that $ \rho_{DM}(t) = \rho_{DM}(0) \; (1+z)^3 $, table I and 
eq.(\ref{rhocrit}). 

We obtain the primordial DM dispersion velocity $ \sigma_{DM} $ 
from eqs. (\ref{roDM2}), (\ref{rhocrit}) and (\ref{F}),
\be\label{sig2}
\sqrt{\frac13 \; \langle \vec{V}^2\rangle(0)} = 
\sigma_{DM} = \left(3 \, \; M_{Pl}^2 \; H_0^2 \; \Omega_{DM} \; \frac1{Z} 
\; \frac{\sigma^3_s}{\rho_s} 
\right)^{\frac13}
\ee
This expression is valid for {\bf any kind} of DM particles.
Inserting eq.(\ref{sig2}) into eq.(\ref{kfs}) yields for the physical free-streaming
wavelength
\be\label{fs2}
\lambda_{fs}(z) = \frac{2 \, \sqrt2 \, \pi}{\Omega_{DM}^\frac16} \;
\left(\frac{3 \; M_{Pl}^2}{H_0}\right)^\frac13 \;
\left(\frac{\sigma_s^3}{Z \; \rho_s}\right)^\frac13 \; \frac1{\sqrt{1+z}} =
\frac{16.3}{Z^\frac13} \;  \; \frac1{\sqrt{1+z}} \; \; {\rm kpc} \; . 
\ee
where we used $ 1 $ keV $ = 1.563738 \; 10^{29} \; ({\rm kpc})^{-1} $.

Notice that $ \lambda_{fs} $ and therefore $ \lambda_J $ turn to be {\bf independent} 
of the nature of the DM particle except for the factor $ Z $.

As stated above $ Z $ for dSphs is in the range:
$$ 
1 < Z < 10000 \; .
$$ 
Therefore, $ 1 < Z^\frac13 < 21.5 $ and the free-streaming wavelength
results in the range
$$
0.757 \; \frac1{\sqrt{1+z}} \; {\rm kpc} <\lambda_{fs}(z) <
16.3 \; \frac1{\sqrt{1+z}} \; {\rm kpc} \; .
$$
These values at $ z = 0 $ are consistent with the $N$-body simulations 
reported in \cite{gao} and are of the order of the small DM structures 
observed today \cite{gilmore}.

The Jeans' mass is given by
\be
M_J(t) = \frac43 \; \pi \; \lambda_J^3(t) \; \rho_{DM} (t) \; .
\ee
and provides the smallest unstable mass by gravitational collapse 
\cite{kt,gb}. Inserting here eq.(\ref{roDM2}) for the DM density and 
eq.(\ref{fs2}) for $ \lambda_J(t) = \lambda_{fs}(t) $ yields 
\be \label{masJ}
M_J(z) = 192 \, \sqrt2 \, \pi^4 \; \sqrt{\Omega_{DM}} \; M_{Pl}^4 \; H_0 \; 
\frac{\sigma_s^3}{Z \; \rho_s} \; (1+z)^\frac32  =
\frac{0.4464}{Z} \; 10^7 \; M_{\odot} \; (1+z)^\frac32 \; .
\ee
Taking into account the $Z$-values range yields
$$
0.4464 \; 10^3 \; M_{\odot} < M_J(z)\; (1+z)^{-\frac32} < 0.4464 \; 10^7  \; 
\; M_{\odot} \; .
$$
This gives masses of the order of galactic masses $ \sim 10^{11} \; 
M_{\odot} $ by the beginning of the MD era $ z \sim 3200 $.
In addition, the comoving free-streaming wavelength scale by 
$ z \sim 3200 $ 
$$
3200 \times \lambda_{fs}(z \sim 3200) \sim 100 \; {\rm kpc}  \; ,
$$
turns to be of the order of the galaxy sizes today.
\subsection{Dark Matter Decoupling at Local Thermal Equilibrium (LTE)}\label{dlte}

If the dark matter particles of mass $ m $ decoupled at a temperature 
$ T_d \gg m $ their freezed-out distribution function only depends on 
$$ 
\frac{p_c}{T_d} = \frac{p_{ph}(t)}{T_d(t)} , \quad {\rm where} \quad
T_d(t) \equiv \frac{T_d}{a(t)} \; .
$$ 
That is, the distribution function for
dark matter particles that decoupled in thermal equilibrium takes the form
$$
F_d^{equil} \left[\frac{p_{ph}(t)}{T_d(t)}\right] = 
F_d^{equil}\left[\frac{p_c}{T_d}\right] \; ,
$$
where $ F_d^{equil} $ is a Bose-Einstein or Fermi-Dirac distribution 
function:
\be \label{fdbe}
F_d^{equil}[p_c] = \frac1{\exp[\sqrt{m^2+p^2_c}/T_d]\pm 1} \; .
\ee
Notice that for eq.(\ref{fdbe}) in this regime:
$$
\frac{\sqrt{m^2+p^2_c}}{T_d} \buildrel{T_d \gg m}\over= y + 
{\cal O}\left(\frac{m^2}{T_d^2}\right) \; .
$$
where $ y $ is defined by  eq.(\ref{varin}) and we can use as distribution 
functions
\be\label{disUR}
F_d^{equil}(y) = \frac1{e^y \pm 1} \; .
\ee
Using eqs.(\ref{m}) and (\ref{fdbe}), we find then for Fermions and 
for Bosons decoupling at LTE
\be\label{mequil}
m = \frac{g_d}{g} \; 
\left\{  \begin{array}{l} 
3.874 \; \mathrm{eV} \; 
~~\mathrm{Fermions} \\
2.906 \; \mathrm{eV} \; ~~\mathrm{Bosons} 
\end{array} \right. \; .
\ee
We see that for DM that decoupled at the Fermi scale: $ T_d \sim 100 $ GeV
and $ g_d \sim 100, \;  m $ results in the keV scale 
as already remarked in ref. \cite{bs,bstpp}.
DM particles may decouple earlier with  $ T_d > 100 $ GeV but
$ g_d $ is always in the hundreds even in grand unified theories 
where $ T_d $ can reach the GUT energy scale.
Therefore, eq.(\ref{mequil}) {\bf strongly suggests} that the mass
of the DM particles which decoupled UR in LTE is in the {\bf keV scale}.

It should be noticed that the Lee-Weinberg \cite{LW} lower bound
as well as the Cowsik-McClelland \cite{cow} upper bound 
follow from eq.(\ref{mdm}) as shown in ref.\cite{oscnos}.

\medskip

Computing the integrals in eq.(\ref{Dex}) with the distribution functions
eq.(\ref{fdbe}) yields for DM decoupling UR in LTE
\be\label{LTDD}
\mathcal{D} = g \; \left\{  \begin{array}{l}
\frac1{4 \; \pi^2} \; \sqrt{\frac{\zeta^5(3)}{15 \; \zeta^3(5)}}
= 1.9625\times 10^{-3}~~\mathrm{Fermions} \\
\frac1{8 \; \pi^2}\; \sqrt{\frac{\zeta^5(3)}{3 \; \zeta^3(5)}}
=3.6569\times 10^{-3}~~\mathrm{Bosons} 
\end{array} \right.
\ee
where $ \zeta(3) = 1.2020569\ldots $ and $ \zeta(5) = 1.0369278 \ldots $.

\medskip

Inserting the distribution function eq.(\ref{disUR}) into eqs.(\ref{mD}) 
and (\ref{gdD}) for $ m $ and $ g_d $, respectively, we obtain 
\be\label{mgdeq}
 m = \left(\frac{Z}{g}\right)^\frac14 \; \mathrm{keV} \; 
\left\{\begin{array}{l}
         0.568~~~\mathrm{Fermions} \\
         0.484~~~\mathrm{Bosons}      \end{array} \right. \quad , \quad
 g_d = g^\frac34 \; Z^\frac14 \; \left\{\begin{array}{l}
         155~~~\mathrm{Fermions} \\
              180~~~\mathrm{Bosons}      \end{array} \right. \; . 
\ee
Since $ g = 1-4 $, for DM particle decoupling at LTE, we see from 
eq.(\ref{mgdeq}) that $ g_d > 100 $ and thus, the DM particle should 
decouple for $ T_d > 100 $ GeV. Notice that $ 1 < Z^\frac14 < 10 $ for 
$ 1 < Z < 10000 $.

\medskip

We can express the free-streaming wavelength as a function of the DM 
particle mass from eqs.(\ref{fs2}) and (\ref{mgdeq}) with the result,
\be\label{fsfdbe}
\lambda_{fs}(z) = \left(\frac{\rm keV}{m}\right)^\frac43 \; 
\frac{\rm kpc}{g^\frac13} \;
\frac1{\sqrt{1+z}} \; \left\{\begin{array}{l}
     7.67 ~~\mathrm{Fermions} \\
     6.19 ~~\mathrm{Bosons} 
 \end{array} \right.  \; .
\ee

\begin{table*}
 \centering
 \begin{minipage}{140mm}
  \begin{tabular}{|c|c|c|} \hline  
Approximation used & Upper limit on $ Z $   &  Upper limit on $ m \simeq 0.5 \; Z^\frac14 \; $ keV \\
\hline 
  Linear fluctuations &  $ \sim 1.3 \times 10^{11} $ & $ 96 $ keV \\
\hline 
Spherical Model &  $ \sim 1.29 \times \delta_i^{-\frac32} \simeq 4.1 \times 10^4 $ & $ 7.1 $ keV\\
\hline   
\end{tabular}
\caption{Upper bounds for the $Z$-factor [defined by eq.(\ref{F1})] and for the mass
of the DM particle obtained for two different approximation methods.
Notice that only the spherical model takes into account non-linear self-gravity effects.
The mass $ m $ {\bf mildly} depends on $ Z $ through the power $ 1/4 $. In any case $ m $
results in the keV range.}
\end{minipage}
\end{table*}

\subsection{The DM particle at the keV scale: conclusions}

Our results on DM are {\bf independent} of the particle model that will 
describe the dark matter. We consider both DM particles that decouple 
being NR and UR and both decoupling at LTE and out of LTE \cite{oscnos,mnras,dsg}.
Our analysis and results refer to the mass of the dark matter particle 
and the number of ultrarelativistic effective degrees of freedom when the
DM particles decoupled. We do not make assumptions about the nature of 
the DM particle and we only assume that its non-gravitational 
interactions can be neglected in the present context (which is
consistent with structure formation and observations).

For DM particles decoupling ultrarelativistic and out of thermal equilibrium
the results for $ m $ and $ g_d $ on the same scales as decoupling at LTE \cite{mnras}.

When DM particles decouple being non-relativistic ($ T_d < m $) the analysis is slightly
more elaborated since it also involves the total annihilation cross-section.
We obtain for non-relativistic DM particles decoupling at LTE \cite{mnras},
\be\label{mtnr}
\sqrt{m \; T_d} = 1.47 \; \left(\frac{Z}{g_d}\right)^\frac13 {\rm keV} \; .
\ee
Therefore, the combination $ \sqrt{m \; T_d} $ must be in the keV scale 
for the NR decoupling case.

In case DM particles explain the formation of galactic
center black holes, DM particles must be fermions with keV-scale mass \cite{mubi}.

\noindent

The mass for the DM particle in the keV range is much larger
than the temperature during the MD era, hence dark matter is {\bf cold} 
(CDM).

A possible CDM candidate in the keV scale is a sterile neutrino \cite{esteril}
produced via their mixing and oscillation with an active 
neutrino species. Other putative CDM candidates in the keV scale are the 
gravitino \cite{gravitino}, the light neutralino  \cite{neutr} 
and the majoron \cite{valle}. 

Actually, many more extensions of the Standard Model
of Particle Physics can be envisaged to
include a DM particle with mass in the keV scale and
weakly enough coupled to the Standard Model particles.

Lyman-$\alpha$ forest observations provide indirect lower bounds on the 
masses of sterile neutrinos \cite{lyalf} while constraints from the 
diffuse X-ray background yield upper bounds on the mass of a putative 
sterile neutrino DM particle \cite{rX}. All these recent constraints are 
consistent with DM particle masses at the keV scale.

The DAMA/LIBRA collaboration has confirmed the presence of a signal in the
keV range \cite{dama}. Whether this signal is due to DM particles
in the keV mass scale is still unclear \cite{contra}. 
On the other hand,
the DAMA/LIBRA signals cannot be explained
by a hypothetical WIMP particle with mass $ \gtrsim O(1) $ GeV since
this would be in conflict with previous WIMPS direct detection experiments 
\cite{wimp}.

\medskip

We find for typical wimps with $ m = 100 $ GeV, $ T_d = 5 $ GeV  \cite{pdg} 
and therefore $ g_d \simeq 80 $ \cite{kt}. This
 requires from eq.(\ref{mtnr}) a huge 
$ Z \sim 10^{23} $, well above the upper bounds  
displayed in Table II \cite{mnras}. Hence, wimps cannot reproduce
the observed galaxy properties. In addition, recall that 
$ Z \sim 10^{23} $ produces
from eq.(\ref{fs2}) an extremely short $ \lambda_{fs} $ today 
$$
\lambda_{fs}(0) \sim 3.51 \; 10^{-4} \; {\rm pc} = 72.4  \; {\rm AU} \; .
$$
No galactic structures has been observed at such solar system scales.

Further evidence for the DM particle mass in the keV scale follows by 
contrasting the observed value of the constant surface density of galaxies 
to the theoretical calculation from the linearized Boltzmann-Vlasov 
equation \cite{dsg}. 

\medskip

In summary, our analysis shows that DM particles decoupling UR in LTE have
a mass $ m $ in the keV scale with $ g_d \gtrsim 150 $  as shown in 
sec. \ref{dlte}. That is, decoupling happens at least at the 100 GeV 
scale. The values of $ m $ and $ g_d $ may be smaller for DM decoupling UR
out of LTE than for decoupling UR in LTE \cite{mnras}. For DM 
particles decoupling NR in LTE we find that 
$ \sqrt{m \; T_d} $ is in the keV range (see eq.(\ref{mtnr}) and \cite{mnras}). 
This is consistent with the DM particle mass in the keV range. 

Notice that the present uncertainity by one order of magnitude
of the observed values of the phase-space density $ \rho_s/\sigma^3_s $ 
only affects the DM particle mass through a power $ 1/4 $ of this 
uncertainity according to eqs. (\ref{mD})-(\ref{mDn}). 
Namely, by a factor $ 10^{\frac14} \simeq 1.8 $.

We find that the free streaming wavelength (Jeans' length) is 
{\bf independent} of the nature of the DM particle except for the $ Z $ 
factor characterizing the decrease of the phase-space density through  
self-gravity [sec. \ref{jeans}]. The values found for the Jeans' length 
and the Jeans' mass for $ m $ in the keV scale 
are consistent with the observed small structure and 
with the masses of the galaxies, respectively.

\medskip

Independent further evidence for the DM particle mass in the keV scale
were recently given in \cite{tikho}. (See also \cite{gilmore}).
DM particles with mass in the keV scale can alleviate CDM problems as the satellite problem 
\cite{sld} and the voids problem \cite{wp}. The DM particle mass in the keV explain
why DM particles were not found in detectors sensitive to particles heavier than $ \sim 1 $ GeV
\cite{CDMS}. In addition, astrophysical mechanisms that can explain the
$ e^+ $ and $ \bar p $ excess in cosmic rays without requiring DM particles
in the GeV scale or above were put forward in \cite{BBS}.

\section{Outlook and future perspectives}

This short review presents the state of the art of the 
effective theory of inflation and 
its successful confrontation with the CMB and LSS data.
We can highlight as perspectives for a foreseeable future:

\begin{itemize}
\item{Measurement of the tensor/scalar ratio $ r $ by CMB
experiments as Planck and the future satellite CMBPol.
This would be the {\bf first} detection of (linearized)
gravitational waves as predicted by Einstein's General Relativity.
In addition, since such primordial gravitational waves were born as quantum
fluctuations, this would be the {\bf first} detection 
of gravitons, namely, {\bf quantized} gravitational waves at tree level.
Such detection of the primordial gravitational waves will test
our prediction $ r \simeq 0.05 $ based on the effective theory of 
slow-roll inflation
(broken symmetric binomial and trinomial potentials) \cite{mcmc,quamc}.}
\item{The running of the spectral index $ dn_s/d \ln k $.
Since the range of the cosmologically relevant modes is
$ \Delta \ln k < 9 $, we have $  \Delta n_s < 9/N^2 \sim 0.0025 $,
where we use the leading order in slow-roll \cite{nos}. Therefore, 
the effective theory of slow-roll inflation indicates that 
the detection of the running calls for measurements of $ n_s $
with a one per thousand precision on a wide range of wavenumbers.}
\item{Non-gaussianity measurements. Although this subject is beyond the 
scope of this short review, let us recall that primordial non-gaussianity is of 
the order $ f_{NL} \sim 1/N $ in single-field slow-roll inflation 
\cite{bartolo}. Such 
small primordial non-gaussianity is hardly expected to be measured in a 
foreseeable future.}
\item{More precise measurements of $ n_s $ together with better data on 
$ r $ and $ dn_s/d \ln k $ will permit to better select the correct
inflationary model. This will test our prediction that a broken symmetric 
inflaton potential with moderate nonlinearity (new inflation) best 
describes the data \cite{nos,mcmc,quamc}.}
\item{Direct dark matter particle detection. Unfortunately, all experiments in course
or planned are built to detect dark matter particles in the GeV scale
or heavier. The same applies for eventual wimps production at the LHC.
A keV scale DM particle will certainly not be detected in such experiments.}
\item{Astrophysical dark matter. Theoretical work, more abundant and better 
data from galaxies and $N$-body computer simulations with keV-scale DM 
will certainly provide relevant new answers to the current problems in
$N$-body computer simulations using heavy ($ m > 1 $ GeV) DM particles.
By current problems we mean the satellite problem, the core-cusp problem,
the void problem and may be the angular momentum problem.
Precise values for $ m $ and $ g_d $ should be obtained.}
\end{itemize}

\end{document}